\documentclass[a4paper,11pt]{article}

\usepackage{jheppub}

\usepackage{amsfonts}
\usepackage{slashed}
\usepackage{dsfont}
\usepackage{graphicx,color}
\usepackage{bbm}

\usepackage{amsmath,amssymb,bbold,epsfig}

\title{\boldmath Quantum field theoretic approach to neutrino oscillations in 
matter}

\author[1]{Evgeny~Kh.~Akhmedov\note{Also at the National Research 
Centre Kurchatov Institute, Moscow, Russia}}

\author{and Alina Wilhelm}

\affiliation{Max-Planck-Institut f\"ur Kernphysik, 
Saupfercheckweg 1 \\  D--69117 Heidelberg, Germany}

\emailAdd{akhmedov@mpi-hd.mpg.de}
\emailAdd{awilhelm@mpi-hd.mpg.de}

\abstract{We consider neutrino oscillations in non-uniform matter in a quantum 
field theoretic (QFT) approach, in which neutrino production, propagation and 
detection are considered as a single process. We find the conditions under 
which the oscillation probability can be sensibly defined and demonstrate 
how the properly normalized oscillation probability can be obtained in the 
QFT framework. We derive the evolution equation for the oscillation 
amplitude and discuss the conditions under which it reduces to the standard 
Schr\"odinger-like evolution equation. It is shown that, contrary to the 
common usage, the Schr\"odinger-like evolution equation is not applicable 
in certain cases, such as oscillations of neutrinos produced in decays of 
free pions provided that sterile neutrinos with $\Delta m^2\gtrsim 1$ eV$^2$ 
exist. }

\begin{document}
\maketitle
\flushbottom



\newcommand{\appsection}{\addtocounter{section}{1}\setcounter{equation}{0}
                         \renewcommand{\thesection}{\Alph{section}}
}

\newcommand{\be}{\begin{equation}}
\newcommand{\ee}{\end{equation}}
\newcommand{\bea}{\begin{eqnarray}}
\newcommand{\eea}{\end{eqnarray}}
%
\renewcommand{\vec}[1]{{\mathbf{#1}}}
%
%

\section{\label{sec:intro}Introduction}

In experiments with solar, atmospheric and supernova neutrinos, the neutrinos 
propagate significant distances in matter before reaching detectors;
the same will also be true for the proposed very long baseline 
accelerator neutrino experiments. 
Matter can affect neutrino oscillations drastically, leading, 
in particular, to their resonance enhancement through the 
Mikheyev-Smirnov-Wolfenstein (MSW) effect \cite{Wolf,MS} or through 
the parametric enhancement mechanism \cite{param}. It is thus extremely 
important to put the analyses of neutrino oscillations in matter on a solid 
theoretical basis. 

The standard approach to neutrino oscillations in matter, pioneered by 
Wolfenstein \cite{Wolf}, is as follows. Mass eigenstate neutrinos $\nu_i$, 
composing a given flavour neutrino state $\nu_\alpha$, are assumed to have 
the same momentum (and, due to their different mass, different energies). 
In the case of oscillations of relativistic neutrinos in vacuum, the 
evolution of the transition amplitude in the flavour basis is then described 
by the Schr\"odinger equation%
\footnote{We use the natural units $\hbar=c=1$ throughout the paper.}   
\be
i\frac{d}{dt}|\nu\rangle = H_0\,|\nu\rangle\,,\qquad~~
H_0 = U\!\left[p\cdot\!{\mathbbm 1}+\frac{M_d^2}{2p} 
\right]\!U^\dag\,.
\label{eq:Sch1}
\ee
Here $|\nu\rangle 
=(|\nu_e\rangle,\,|\nu_\mu\rangle,\,|\nu_\tau,\rangle\,...)^T$ is the 
flavour-basis neutrino state vector (with dots standing for possible 
light sterile neutrino states), 
$p$ is the neutrino momentum, $M_d={\rm diag} (m_1, \, m_2, \, m_3,\,...)$ 
is the neutrino mass matrix in the mass eigenstate basis, and $U$ is the 
leptonic mixing matrix which relates the flavour-eigenstate neutrino state 
vectors $|\nu_\alpha\rangle$ ($\alpha=e,\mu,\tau, \dots$) with the 
mass-eigenstate ones $|\nu_i\rangle$ ($i=1, 2, 3, ...)$:
\be
|\nu_\alpha\rangle = \sum_i U^*_{\alpha i}|\nu_i\rangle\,.
\label{eq:mix1}
\ee
Since the first term in the square brackets in the expression for 
$H_0$ in (\ref{eq:Sch1}) is proportional to the unit matrix, it leads to a 
common phase shift of all neutrino flavour states and therefore does not 
affect the oscillation probabilities. Thus, it can be omitted from $H_0$. The 
matter effects on neutrino oscillations can be incorporated by replacing the 
free Hamiltonian $H_0$ by the effective Hamiltonian of neutrino propagation in 
matter according to $H_0\to H=H_0+V$. Here $V$ is the matrix of matter-induced 
neutrino potentials due to coherent forward scattering of neutrinos on 
matter constituents; in the absence of background neutrinos it is diagonal 
in the flavour-eigenstate basis. Taking into account that for pointlike 
relativistic neutrinos the distance $x$ they propagate over the time $t$ 
satisfies $x\simeq t$ and that to leading order in small neutrino masses 
$p\simeq E$ where $E$ is the average neutrino energy, one arrives at the 
following equation describing neutrino flavour evolution in matter: 
\be
i\frac{d}{dx}|\nu\rangle = \left[U\frac{M_d^2}{2E} U^\dag + V(x)\right]
|\nu\rangle\,.
\label{eq:Sch2}
\ee

Equation~(\ref{eq:Sch2}) is employed in virtually all studies of neutrino 
oscillations in non-uniform matter. However, its derivation presented above 
was based on heuristic considerations and it certainly needs to be put on a 
more solid ground. 
Attempts at deriving eq.~(\ref{eq:Sch2}) within the relativistic quantum 
mechanics and quantum field theory (QFT) frameworks have been made in a number 
of papers. In \cite{Halprin} evolution of Dirac neutrinos in matter was 
described by making use of a Dirac equation with matter-induced potential. It 
was demonstrated that the neutrino wave function satisfies eq.~(\ref{eq:Sch2}) 
provided that matter density varies little over the distances of order of the  
neutrino de~Broglie wavelength. The Dirac equation was also employed in 
\cite{changzia}, though only the case of matter of constant density 
was considered there. In ref.~\cite{mannheim} the Dirac equation was used 
to describe the evolution of Dirac and Majorana neutrinos in matter, but 
again only in the case of constant-density matter. Besides the already 
mentioned ref.~\cite{Halprin}, the Dirac equation has been employed for 
describing neutrino oscillations in non-uniform matter in 
refs.~\cite{sawyer,grimus}, both in the Dirac \cite{sawyer,grimus} and 
Majorana \cite{grimus} neutrino cases.  
In none of these papers, however, 
neutrino production and detection processes were included in the 
description of neutrino oscillations. The most advanced study of neutrino 
evolution in matter of varying density was carried out in \cite{cardchung} 
in the QFT framework. In that paper the treatment included the neutrino 
production and detection processes,  
and it was also demonstrated 
how the correctly normalized oscillation probability can be obtained. 
The employed normalization procedure was rather cumbersome, though. 

The main goal of refs. [4-9] was to derive the evolution 
equation~(\ref{eq:Sch2}) 
from relativistic quantum mechanics or QFT. However, the conditions under 
which this equation is valid were not fully studied in those papers.
Furthermore, the question of how neutrino oscillations can be described 
in the situations when these conditions are not satisfied (and consequently 
eq.~(\ref{eq:Sch2}) cannot be used) was not addressed. In addition, no 
discussion of neutrino production and detection coherence and of their effect 
on neutrino oscillations in matter was given. 

\begin{figure}
  \begin{center}
    \includegraphics{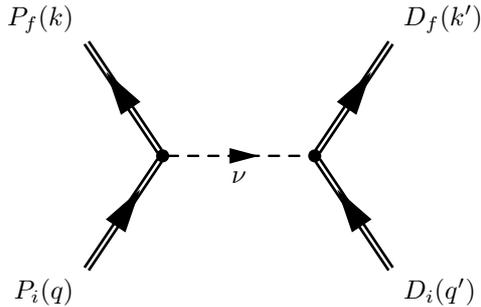}
  \end{center}
\caption{Feynman diagram describing neutrino production, propagation
           and detection as a single process.}
  \label{fig:feyn}
\end{figure}

In the present paper we consider neutrino oscillations in non-uniform matter 
in the framework of QFT. In this approach neutrino production, propagation, 
and detection are treated as a single process, described by a Feynman diagram 
with the neutrino in the intermediate state (such as the one in 
fig.~\ref{fig:feyn}). We discuss the conditions under which the oscillation 
probability can be extracted from the rate of the overall neutrino 
production-propagation-detection process and demonstrate that this probability 
is automatically correctly normalized and satisfies the unitarity constraints. 
We also identify the conditions under which the amplitude of neutrino flavour 
transition can be found as a solution of eq.~(\ref{eq:Sch2}). One of our main 
results is that eq.~(\ref{eq:Sch2}) is {\em not} applicable when neutrino 
production and/or detection coherence is violated. We discuss the situations 
when this can happen and consider an alternative way of describing neutrino 
oscillations in those cases.  

Our treatment of neutrino oscillations in non-uniform matter closely 
parallels the treatment of neutrino oscillations in vacuum performed in 
ref.~\cite{AK}, but differs from the latter in a number of important 
aspects. The differences are mostly related to the properties of the neutrino 
propagator in non-uniform matter, which deviate significantly from those of 
the vacuum neutrino propagator and do not allow using some techniques that 
were applied to the neutrino propagator in vacuum.   
To make it easier to follow our treatment, let us briefly outline our main 
steps.
\begin{itemize} 

\item
We consider the neutrino production, propagation and detection process 
described by the Feynman diagram of fig.~\ref{fig:feyn}. The external 
legs in this diagram correspond to  
the particles that accompany neutrino production and detection. These are 
either propagating particles or bound states, which are described by the 
suitable state vectors. The intermediate neutrino state is described by 
a propagator, which is found as a solution of the corresponding Dirac equation 
with matter-induced potential for neutrinos $V(\vec{x})$. 

\item
Since this potential depends on the coordinate $\vec{x}$, the system is not 
translationally invariant and the neutrino momentum is not conserved. 
As a result, the neutrino propagator in the momentum space 
depends on two momenta, $\vec{p}$ and $\vec{p}'$, rather than one. 
The amplitude of the overall process can be written as the integral over these 
two momenta, with the production and detection amplitudes $\Phi_P(\vec{p})$ and 
$\Phi_D(\vec{p}')$ multiplying the momentum-space neutrino propagator 
$\tilde{S}(E; \vec{p'}, \vec{p})$ in the 
integrand. 

\item
For the intervals of momenta over which 
the propagator varies significantly, the neutrino production and detection 
amplitudes $\Phi_P(\vec{p})$ and $\Phi_D(\vec{p}')$ change very little.
This allows one to greatly simplify the expression for the amplitude of 
the process. 

\item
We calculate the rate of the overall 
neutrino production-propagation-detection process and identify the 
conditions under which this rate factorizes into the neutrino production 
rate $d\Gamma_\alpha^{prod}/dE$, propagation (oscillation) probability 
$P_{\alpha\beta}$ and the detection cross section $\sigma_\beta$. When 
these conditions are satisfied, the oscillation probability can be 
extracted from the rate of the overall process by dividing the latter by 
$d\Gamma_\alpha^{prod}/dE$ and $\sigma_\beta$. 

\item
We reconstruct the oscillation amplitude from the expression for 
$P_{\alpha\beta}$ and derive the equation that it obeys. We identify the 
conditions under which this equation coincides with eq.~(\ref{eq:Sch2}) and 
also discuss the situations when these conditions are violated and 
eq.~(\ref{eq:Sch2}) is not applicable. 
\end{itemize} 

The described above program is realized in sections 
\ref{sec:ampl}-\ref{sec:evoleq} of the paper; in section \ref{sec:disc} we 
summarize and discuss the obtained results. To make the paper self-contained, 
in Appendix A we briefly review the derivation of the neutrino propagator in 
non-uniform matter performed in \cite{cardchung}, whereas in Appendix B we 
give a compendium of expressions for neutrino potentials in matter. 
Appendices~C and~D contain  
derivation of some results used in sections~\ref{sec:simpl} 
and~\ref{sec:evoleq}. 

\section{\label{sec:ampl}The transition amplitude}
\subsection{\label{sec:gen} General formalism}

Consider the process of neutrino production, propagation and detection 
described by the Feynman diagram of fig.~\ref{fig:feyn}. We shall be assuming 
that the neutrino production process involves one initial state and one final 
state particle (besides the neutrino). Likewise, the detection process 
will also be assumed to involve only one particle besides the neutrino in 
the initial state and one particle in the final state. The generalization to 
the case of an arbitrary number of particles involved in the neutrino 
production and detection processes is straightforward and would just 
make the formalism more cumbersome without providing further physical 
insight.%
\footnote{
As only one particle is assumed to be in the initial state of the 
production process, it must be unstable. This will be of no importance for us 
here, though.}

Let us first discuss the state vectors of the particles accompanying neutrino 
production and detection (external particles). In quantum theory, 
one-particle states of particles of type $A$ can be written as   
\be
|A\rangle =\int\! 
[d p]
\,f_A(\vec{p},
\vec{P})\,|A,\vec{p}
\rangle\,,
\label{eq:state1}
\ee
where $|A,\vec{p}\rangle$ is the one-particle momentum eigenstate  
corresponding to momentum $\vec{p}$ and energy $E_A(\vec{p})$,  
and  $f_A(\vec{p},\vec{P})$ is the momentum distribution 
function with the mean momentum $\vec{P}$. In eq.~(\ref{eq:state1}) we use 
the shorthand notation 
\be
[dp]\equiv\frac{d^3 p}{(2\pi)^3\sqrt{2E_A(\vec{p})}}\,.
\label{eq:not1}
\ee
For particles with spin, the states $|A\rangle$ and $|A,\vec{p}\rangle$ 
depend also on a spin variable, which we suppress to simplify the notation. 

Throughout this paper we will be using the normalization conventions of 
ref.~\cite{PeskSchr} and the notation $P_{L,R}=(1\mp\gamma_5)/2$.  
We choose the Lorentz invariant normalization condition for the plane 
wave states $|A,\vec{p}\rangle$:\be
\langle 
A,\vec{p}'|A,\vec{p}\rangle= 2E_A(\vec{p})\,(2\pi)^3 
\delta^{(3)}(\vec{p}-\vec{p'})\,. 
\label{eq:norm1}
\ee
The standard normalization of the states $\langle A|A\rangle=1$ then implies 
\be
\int\! \frac{d^3p}{(2\pi)^3}\,|f_A(\vec{p})|^2=1\,. 
\label{eq:norm2} 
\ee

The states describing the external particles in fig.~\ref{fig:feyn} can be 
represented in the form (\ref{eq:state1}). For the initial and final states at 
neutrino production we write
\be
|P_i\rangle =\int\! [d q] 
\,f_{Pi}(\vec{q},\vec{Q})\,|P_i,\vec{q}\rangle\,,\qquad
|P_f\rangle =\int\! [d k]
\,f_{Pf}(\vec{k},\vec{K})\,|P_f,\vec{k}\rangle\,,
\label{eq:state2}
\ee
and similarly for the states participating in neutrino detection: 
\be
|D_i\rangle =\int\! [d q']
\,f_{Di}(\vec{q}',\vec{Q}')\,|D_i,\vec{q}'\rangle\,,\qquad
|D_f\rangle =\int\! [dk']
\,f_{Df}(\vec{k}',\vec{K}')\,|D_f,\vec{k}'\rangle\,.
\label{eq:state3}
\ee
We assume these states to obey the normalization condition 
\eqref{eq:norm2}. Some (or all) of the mean momenta of the external particles
$\vec{Q}$, $\vec{K}$, $\vec{Q}'$ and $\vec{K}'$ may vanish, i.e.\ the states 
in eqs.~(\ref{eq:state2}) and~(\ref{eq:state3}) can describe bound states 
at rest as well as wave packets.

The amplitude of the neutrino production - propagation - detection process is 
given by the matrix element
\be
i{\cal A}_{\beta\alpha}=\langle P_f \,D_f|\hat{T}\exp\Big[-i\int \!d^4x 
\,{\cal H}_I(x)\Big]-\mathbbm{1}|P_i \,D_i\rangle\,,
\label{eq:amp2}
\ee
where $\hat{T}$ is the time ordering operator and ${\cal H}_I(x)$ is the 
charged-current weak interaction Hamiltonian. 
{}From this equation it is easy to calculate the transition amplitude 
in the lowest nontrivial order in ${\cal H}_I$ using the standard QFT methods. 
The resulting expression corresponds to the Feynman diagram of 
fig.~\ref{fig:feyn} and can be written as 
\bea
i{\cal A}_{\beta\alpha}&=&
\int\! [dq]\,f_{Pi}(\vec{q},\vec{Q}) \int\! [dk]
\,f^*_{Pf}(\vec{k},\vec{K}) 
\nonumber \\
& &
\times \int\! [dq']
\, f_{Di}(\vec{q}',\vec{Q}') \int\! [dk']
\,f^*_{Df}(\vec{k}',\vec{K}')
\,i{\cal A}^{p.w.}_{\beta\alpha}(q,k;q',k')\,.
\label{eq:amp3}  
\eea
Here 
the quantity ${\cal A}^{p.w.}_{\beta\alpha}(q,k;q',k')$ is the 
amplitude of the process with plane-wave external states:
\begin{align}
i{\cal A}^{p.w.}_{\beta\alpha}
(q,k;q',k')=
\int d^4 x_1 \!\int d^4 x_2 \,
&\tilde{M}_D(q',k')\, e^{-i(q'-k')(x_2-x_D)}
\nonumber \\
& \times
S_{\beta\alpha}(x_2, x_1)
\tilde{M}_P(q,k)\, e^{-i(q-k)(x_1-x_P)} \,.\qquad 
\label{eq:amp4}
\end{align}
In this equation $x_1$ and $x_2$ are the 4-coordinates of the neutrino 
production and detection points.  
The choice of the \mbox{4-coordinate} dependent phase factors 
corresponds 
to the assumption that the peaks of the wave packets of particles involved in 
the production process are all located at $\vec{x}_1=\vec{x}_P$ at the time 
$t_1=t_P$, whereas for the detection process the corresponding peaks are all 
situated at $\vec{x}_2=\vec{x}_D$ at the time $t_2=t_D$ (this assumption can 
be relaxed, see section 6.2 of ref.~\cite{AK}). 
The quantities $\tilde{M}_P(q,k)$ and 
$\tilde{M}_D(q',k')$ are the plane-wave amplitudes of the processes 
$P_i\to P_f+\nu_\alpha$ and $D_i+\nu_\beta\to D_f$,
respectively, with the neutrino spinors excluded. 
They are related to the full plane-wave neutrino production and detection 
amplitudes 
$M_P(q,k)$ and $M_D(q',k')$ through 
\be
M_{P}(q,k) = \frac{\bar{u}_{L}(p)}{\sqrt{2 p_0}}\tilde{M}_P(q,k)\,\qquad 
\mbox{and}\qquad M_{D}(q',k') = \tilde{M}_D(q',k')\frac{u_{L}(p')}
{\sqrt{2 p_0'}}\,.
\label{eq:M}
\ee
Here $u_L(p)$ is the left-handed neutrino spinor,%
\footnote{We do not write flavour 
indices for the neutrino spinors because in the limit of ultra-relativistic 
neutrinos that we consider the spinors $u_L(p)$ correspond to 
essentially massless neutrinos.}
$p=q-k$, $p'=q'-k'$, and 
$p_0,\, p_0'$ are the time components of the corresponding 4-momenta.

The quantity $S_{\beta\alpha}(x_2,x_1)$ in the 
second line of eq.~(\ref{eq:amp4}) is the coordinate-space neutrino 
propagator in matter in the flavour basis. It is a matrix in both 
flavour space and spinor space, whereas the quantities $\tilde{M}_P(q,k)$ 
and $\tilde{M}_D(q',k')$ are Dirac spinors. To simplify the notation, we have 
suppressed the corresponding spinor indices. 

The neutrino propagator in matter $S_{\beta\alpha}(x,x')$ 
satisfies the Schwinger-Dyson equation 
\be
[i\gamma_\mu \partial^\mu
-M P_R-M^\dag P_L] 
S(x, x')-\int d^4 x''\, \Sigma(x, x'') P S(x'', x')=
\delta^4(x-x')\!\cdot\! \mathbbm{1}\,,
\label{eq:DirProp}
\ee
where $M$ is the neutrino mass matrix in the flavour basis, $\Sigma(x,x')$ is 
the matter-induced neutrino self-energy, $\mathbbm{1}$ is the unit matrix in 
the flavour space, and flavour indices are suppressed for simplicity. 
Eq.~(\ref{eq:DirProp}) (as well as eq.~(\ref{eq:Dir1}) below) applies to 
both Dirac and Majorana neutrino cases, provided that for Majorana neutrinos 
one uses the Feynman rules with propagators and vertices not containing 
explicitly the charge-conjugation matrix (see, e.g., \cite{gates,denner}). 
The operator $P$ in (\ref{eq:DirProp}) is defined as $P=P_L$ for Dirac 
neutrinos and $P=-\gamma_5$ for Majorana neutrinos. We will discuss the 
choice of the operator $P$ in the Majorana neutrino case 
in Appendix A. 

For Dirac neutrinos,  
the mass matrix $M$ is a general $N_f\times N_f$ matrix where $N_f$ is 
the number of light neutrino species. Note that for $M\ne M^\dag$ the presence 
of the term $M P_R+M^\dag P_L$ rather than the usual mass term $M$ in 
(\ref{eq:DirProp}) is required by the hermiticity of the Lagrangian.
For Majorana neutrinos, $M=M^T$. 

The matter-induced neutrino self-energy $\Sigma(x,x')$ is due to neutrino 
interaction with the particles of the medium through the exchange of $W^\pm$ 
and $Z^0$ bosons. For low energies of neutrinos and background particles, one 
can expand the propagators of the $W^\pm$ and $Z^0$ bosons in the inverse 
powers of their squared mass; the leading (neutrino energy and momentum 
independent) terms in these expansions yield 
\be
\Sigma(x, x')\simeq \Sigma_0(x)\delta^4(x-x')\,,\quad \mbox{where}\qquad
\Sigma_0(x)=\gamma_\mu V^\mu(x)\,,
\label{eq:approx}
\ee
and $V^\mu(x)$ can be considered as an effective neutrino potential. Next 
(finite-order) terms in the expansions in $1/m_{W}^2$ and $1/m_{Z}^2$ bring in 
some neutrino momentum dependence, which in the coordinate representation would 
result in the appearance of derivative terms in eq.~(\ref{eq:approx}); however, 
upon integration over $x''$ in eq.~(\ref{eq:DirProp}) these terms would still 
lead to local terms in the self-energy $\Sigma$. Thus, in the case when only a 
finite number of terms in the expansion of the propagators of the intermediate 
bosons in inverse powers of their squared mass is kept, neutrino interaction 
with matter can be described by a local effective potential $V(x)$.%
\footnote{Note that the term `potential' is not very precise. Strictly 
speaking it applies solely to the case when only leading terms of expansions 
in powers of $1/m_W^2$ and $1/m_Z^2$ are retained. Otherwise, $V(x)$ would 
depend on neutrino energy. This will pose no problem if $V(x)$ is 
time-independent (i.e.\ $V(x)=V(\vec{x})$), so that different neutrino energy 
modes can be studied separately. It would be more correct to call $V(x)$ 
local matter-induced neutrino self energy. We use the term `effective 
potential' for brevity.}
Eq.~(\ref{eq:DirProp}) can then be rewritten as  
\be
\big[\gamma_\mu (i\partial^\mu-V^\mu(x) P)\delta_{\beta\gamma}
-M_{\beta\gamma}P_R-M^\dag_{\beta\gamma}P_L\big] S_{\gamma\alpha}(x, x')=
\delta^4(x-x')\delta_{\alpha\beta}\,,
\label{eq:Dir1}
\ee
\noindent
where we have reinstated the flavour indices, 
whereas spinor indices are suppressed as before. For a non-relativistic medium 
of unpolarized particles, only the time component of $V^\mu(x)$ is essentially 
non-zero: $V_\mu(x)\simeq V(x)\delta_{\mu 0}$. 
The solution of eq.~(\ref{eq:Dir1}) for the neutrino propagator in the case 
of Dirac neutrinos was given in \cite{cardchung}. It is reviewed in 
Appendix~A, 
where also the Majorana neutrino case is considered.  

Let us now discuss the spinor structure of the neutrino propagator and 
of the plane-wave  
production and detection amplitudes. The 
propagator can be  represented as 
\be
S=\left(
\begin{array}{cc}
S_{LL} & S_{LR} \\
S_{RL} & S_{RR} 
\end{array}
\right),
\label{eq:S}
\ee
where 
$S_{LL}=P_L S P_L$, $S_{LR}=P_L S P_R$, etc., are $2\times 2$ 
block matrices, 
and we have now omitted the flavour indices. 
Since only left-handed neutrinos participate in weak interactions, we are 
interested only in the $LR$-component of the propagator: $-i\langle 
\hat{T}\nu_L(x)\bar{\nu}_L(x')\rangle=P_L 
S(x, x') P_R=S_{LR}(x,x')$. It will be convenient for us to work in the 
chiral (i.e.\ Weyl) representation of the Dirac \mbox{$\gamma$-matrices}, in 
which $\gamma_5$ is diagonal. It can then be shown that for ultra-relativistic 
neutrinos with spin down along the 3rd spatial axis only the 
\mbox{22-component} 
of the $2\times 2$ matrix $S_{LR}(x, x')$ is non-zero \cite{cardchung}. 
Likewise, from the Dirac equation it follows that the left-handed 2-component 
neutrino spinors in the momentum space are $u_L(p)=(0, \sqrt{2 p_0})^T$ 
(see, e.g., \cite{PeskSchr}, eq.~(3.53)). 
{}From eq.~(\ref{eq:M}) we then find
\be
M_{P}(q,k) = \tilde{M}_{P2}(q,k)\,\qquad 
\mbox{and}\qquad M_{D}(q',k')=\tilde{M}_{D2}(q',k')\,,
\label{eq:M2}
\ee
where the index 2 stands for the second (lower) components of the 
left-handed spinors $\tilde{M}_{P}(q,k)$ and $\tilde{M}_{D}(q',k')$. 
On the other hand, we have 
\be
\tilde{M}_{D}S_{LR}\tilde{M}_{D}=
\tilde{M}_{D2}(S_{LR})_{22}\tilde{M}_{P2}\,,
\label{eq:rel1}
\ee
where we have taken into account that only the 22-component of $S_{LR}$ 
is different from zero. Denoting this component as $\hat{S}$, from 
(\ref{eq:M2}) and (\ref{eq:rel1}) we find  
\be
\tilde{M}_{D}S_{LR}\tilde{M}_{D}=M_{D}\,\hat{S}\,M_{P}\,.
\label{eq:rel2}
\ee
We can now rewrite eq.~(\ref{eq:amp4}) as 
\begin{align}
i{\cal A}^{p.w.}_{\beta\gamma}
(q,k;q',k')=
\int d^4 x_1 \!\int d^4 x_2 \,
&M_D(q',k')\, e^{-i(q'-k')(x_2-x_D)}
\nonumber \\
& \times
\hat{S}_{\beta\gamma}(x_2, x_1)
M_P(q,k)\, e^{-i(q-k)(x_1-x_P)} \,.\qquad 
\label{eq:amp5}
\end{align}
where the integrand does not contain any quantities with spinor indices.   

It will be convenient for us to express the coordinate-space neutrino 
propagator $\hat{S}_{\beta\alpha}(x_2, x_1)$ as a Fourier transform of 
the momentum-space one. We will be assuming that the matter-induced potential 
of neutrinos $V^\mu$ depends on the spatial coordinate $\vec{x}$ 
but is time-independent: $V^\mu=\hat{V}^\mu(\vec{x})$.%
\footnote{This is a good approximation provided that the potential is nearly  
static over the time intervals of order 
$\Delta t=\sigma_{x\nu}/v_\nu$, where $\sigma_{x\nu}$ is the length 
of the neutrino wave packet and $v_\nu\approx 1$ is the neutrino velocity.} 
In this case the system under consideration possesses translation invariance 
in time but not in space; as a result, the coordinate-space propagator depends 
on the times $t_1$ and $t_2$ only through their difference, but on the spatial 
coordinates $\vec{x}_1$ and $\vec{x}_2$ separately: $\hat{S}(x_2, x_1)=
\hat{S}(t_2-t_1;\vec{x}_2, \vec{x}_1)$. This constitutes an important 
difference as compared to the case of neutrino propagation in vacuum, where 
the coordinate-space neutrino propagator depends only on $x_2-x_1=(t_2-t_1;\, 
\vec{x}_2-\vec{x}_1)$. The momentum-space neutrino propagator in 
non-uniform but static matter will therefore depend on one energy 
variable $p_{0}=p_{0}'\equiv E$ and two momenta: 
$\tilde{S}=\tilde{S}(E;\vec{p}', \vec{p})$. The coordinate-space neutrino 
propagator is related to the momentum-space one through the Fourier 
transformations with respect to the energy and both momenta: 
\be
\hat{S}_{\beta\alpha}(t_2-t_1; \vec{x}_2, \vec{x}_1)=
\int\frac{dE}{2\pi}
\frac{d^3 p}{(2\pi)^3}\frac{d^3 p'}{(2\pi)^3}\,
\tilde{S}_{\beta\alpha}(E;\vec{p}', \vec{p})\,
e^{-iE(t_2-t_1)}\,e^{i\vec{p}'\vec{x}_2-i\vec{p}\vec{x}_1}\,. 
\label{eq:prop1}
\ee
Substituting this into eq.~(\ref{eq:amp4}), going to the shifted 
integration variables $x_1'=x_1-x_P$ and $x_2'=x_2-x_D$ and 
then using the obtained result in eq.~(\ref{eq:amp3}), we find   
\be
{\cal A}_{\beta\alpha}=
 \int\frac{dE}{2\pi}
\frac{d^3 p}{(2\pi)^3}\frac{d^3 p'}{(2\pi)^3}\,\Phi_{D}(E,\vec{p}')\,
\tilde{S}_{\beta\alpha}(E;\vec{p}', \vec{p})\,
\Phi_{P}(E, \vec{p})\,
e^{-iE(t_D-t_P)}\,e^{i\vec{p}'\vec{x}_D-i\vec{p}\vec{x}_P}\,. 
\label{eq:amp6}
\ee
Here the functions $\Phi_{P}(E, \vec{p})$ and $\Phi_{D}(E, \vec{p}')$ are 
given by 
\begin{align}
\Phi_{P}(E, \vec{p})\,=&\int d^4 x_1' e^{i p x_1'}\int\!
[d q]\int\! [dk]
\,f_{Pi}(\vec{q},\vec{Q})\,f^*_{Pf}(\vec{k},\vec{K}) \,
e^{-i(q-k)x_1'}M_{P}(q,k)\,, 
\label{eq:psi} \\
\Phi_{D}(E, \vec{p}')\,=&\int d^4 x_2' e^{-i p' x_2'}\int\!
[dq'] \int\! [dk']
\,f_{Di}(\vec{q}',\vec{Q}')\,f_{Df}^*(\vec{k}',\vec{K}') \,e^{-i(q'-k')x_2'}
M_{D}(q',k')\,,
\nonumber
\label{eq:Phi}
\end{align}
where the 4-vectors $p$ and $p'$ are defined as $p=(E, \vec{p})$, 
$p'=(E, \vec{p}')$. 

The quantities $\Phi_{P}(E, \vec{p})$ and $\Phi_{D}(E, \vec{p}')$ are the 
amplitudes of the neutrino production and 
detection processes in which the external particles are described by the 
state vectors (\ref{eq:state2}) and (\ref{eq:state3}), while the produced 
and detected neutrino states are described by plane waves of \mbox{4-momenta} 
$p$ and $p'$, respectively. 
They are thus the probability amplitudes that the 
emitted and detected neutrinos have the corresponding 4-momenta, i.e. 
are the amplitudes of the momentum distribution functions of these 
neutrinos. In the limit of plane-wave external particles, 
$\Phi_{P}(E, \vec{p})$ and $\Phi_{D}(E, \vec{p}')$ are proportional to 
$\delta$-functions expressing the momentum conservation at neutrino 
production and detection. In the realistic case when the external particles 
are described by wave packets, the functions $\Phi_{P}(E, \vec{p})$ and 
$\Phi_{D}(E, \vec{p}')$ represent approximate conservation of mean momenta 
in the neutrino production and detection processes and are characterized by 
peaks of finite widths, 
with peak momenta being, respectively, $\vec{P}\approx\vec{Q}-\vec{K}$ 
and $\vec{P}'\approx\vec{Q}'-\vec{K}'$ \cite{AK}. 
The width $\sigma_{pP}$ of the peak of the function $\Phi_{P}(E, \vec{p})$ 
depends on the momentum uncertainties of the particles taking part in 
neutrino production. It is dominated by the largest of these uncertainties:
$\sigma_{pP}\sim \max\{\sigma_{Pi}, \sigma_{Pf}\}$. Quite analogously, 
the width $\sigma_{pD}$ of the peak of the function $\Phi_{D}(E, \vec{p})$ 
satisfies $\sigma_{pD}\sim \max\{\sigma_{Di}, \sigma_{Df}\}$. 

By the Heisenberg uncertainty relations, the momentum uncertainties at 
neutrino production and detection are related to the spatial localizations 
of the neutrino production and detection processes, $\sigma_{xP}$ and 
$\sigma_{xD}$: 
\be 
\sigma_{pP}\sim \frac{1}{\sigma_{xP}}\,, \qquad
\sigma_{pD}\sim \frac{1}{\sigma_{xD}}\,.
\label{eq:uncert1}
\ee

\subsection{\label{sec:simpl} Transition amplitude: a simplification}

Let us now proceed with the calculation of the transition amplitude. 
In the case of neutrino oscillations in vacuum, there exists a closed-form 
expression for the neutrino propagator, which in the momentum space depends 
on just one 4-momentum $p$. For the coordinate-space propagator one can 
then use an asymptotic expression at large baselines $L$ given by the 
so-called Grimus-Stockinger theorem \cite{Grimus:1996av}. This  leads to a 
considerable simplification of the expression for the amplitude of the overall 
neutrino production-propagation-detection process \cite{AK}. Unfortunately, 
for neutrino oscillations in matter with an arbitrary density profile no 
closed-form expression for the neutrino propagator exists, and the 
Grimus-Stockinger theorem cannot be utilized. One therefore has to find 
another way to proceed with the computation. We shall show now that 
the calculations can be greatly simplified by making use of the 
fact that the momentum dependence of different factors in the integrand in 
eq.~(\ref{eq:amp6}) has different character.  

Let us first note that the phase $\vec{p}'\vec{x}_D-\vec{p}\vec{x}_P$ of the 
momentum-dependent complex phase factor in~(\ref{eq:amp6}) can be written as 
$\frac{1}{2}(\vec{p}'-\vec{p})(\vec{x}_D+\vec{x}_P)+\frac{1}{2}(\vec{p}'
+\vec{p})(\vec{x}_D-\vec{x}_P)$. The first term here can be eliminated by 
the proper choice of the origin of the coordinate frame. The second term  
implies that the exponential factor in the integrand of~(\ref{eq:amp6})
varies significantly when the momenta 
$\vec{p}, \vec{p}'$ vary by $|\Delta\vec{p}|, |\Delta\vec{p}'|\sim L^{-1}$, 
where $L=|\vec{x}_D-\vec{x}_P|$ is the baseline. Since $L$ is a macroscopic 
distance, 
the phase factor in the integrand of 
(\ref{eq:amp6}) is a fast oscillating function of the momenta $\vec{p}$ and 
$\vec{p}'$. At the same time, the neutrino production and detection amplitudes 
$\Phi_{P}(E, \vec{p})$ and $\Phi_{D}(E, \vec{p}')$ are slowly varying 
functions of the momenta. Indeed, they change significantly when the 
corresponding momenta vary by $|\Delta\vec{p}|\sim\sigma_{pP}\sim 
1/\sigma_{xP}$ and $|\Delta\vec{p}'|\sim \sigma_{pD}\sim 1/\sigma_{xD}$. 
Because the sizes of localization regions of the neutrino production and 
detection regions are by far much smaller than the oscillation baselines of 
interest,%
\footnote{\label{foot:f1}For instance, if $\sigma_{xP}$ and $\sigma_{xD}$ are 
of the order of interatomic distances and $L\sim 1$ km, then $\{\sigma_{xP},
\sigma_{xD}\}/L\sim 10^{-13}$.}
the amplitudes  $\Phi_{P}(E, \vec{p})$ and 
$\Phi_{D}(E, \vec{p}')$ change very little over the momentum intervals over 
which the phase factor varies significantly. Therefore, these amplitudes  
can be pulled out of the momentum integrals at the values of momenta 
$\vec{p}=\vec{p}_*$ and $\vec{p}'=\vec{p}'_*$, where $\vec{p}_*$ and 
$\vec{p}'_*$ are the central momenta of the regions which give the main 
contributions to the integrals over $\vec{p}$ and $\vec{p}'$, respectively. 
As a result, eq.~(\ref{eq:amp6}) becomes 
\be
{\cal A}_{\beta\alpha}=
\int\frac{dE}{2\pi}\Phi_{D}(E,\vec{p}_*')\,\Phi_{P}(E, \vec{p}_*)\,
e^{-iE(t_D-t_P)}
\int \frac{d^3 p}{(2\pi)^3}\frac{d^3 p'}{(2\pi)^3}\,
\tilde{S}_{\beta\alpha}(E;\vec{p}', \vec{p})\,
e^{i\vec{p}'\vec{x}_D-i\vec{p}\vec{x}_P}\,.
\label{eq:amp7}
\ee
The last integral here is nothing but the neutrino propagator in the mixed 
energy-coordinate representation: 
\be
\hat{S}_{\beta\alpha}(E;\vec{x}_D, \vec{x}_P)\equiv\int\,d\tau
e^{i E \tau} \hat{S}_{\beta\alpha}(\tau; \vec{x}_D, \vec{x}_P)=
\int \frac{d^3 p}{(2\pi)^3}\frac{d^3 p'}{(2\pi)^3}\,
\tilde{S}_{\beta\alpha}(E;\vec{p}', \vec{p})\,
e^{i\vec{p}'\vec{x}_D-i\vec{p}\vec{x}_P}\,, 
\label{eq:S2}
\ee
where the last equality follows from eq.~(\ref{eq:prop1}). Thus, we finally 
obtain 
\be
{\cal A}_{\beta\alpha}=
\int\frac{dE}{2\pi}\Phi_{D}(E,\vec{p}_*')\,\Phi_{P}(E, \vec{p}_*)\,
e^{-iE(t_D-t_P)}\,
\hat{S}_{\beta\alpha}(E;\vec{x}_D, \vec{x}_P)\,. 
\label{eq:amp8}
\ee

Let us now discuss the propagator $\hat{S}_{\beta\alpha}(E;\vec{x}, 
\vec{x}')$. It has been shown in \cite{cardchung} (see also Appendix A) 
that this quantity can be represented as 
\be 
\hat{S}_{\beta\alpha}(E; \vec{x},\vec{x}')=-2E\frac{e^{i|E||\vec{x}-
\vec{x}'|}}{4\pi|\vec{x}-\vec{x}'|}
\hat{F}_{\beta\alpha}(E; \vec{x},\vec{x}')\,,
\label{eq:propF1}
\ee
where $E>0$ for neutrinos, $E<0$ for antineutrinos, and 
$F_{\beta\alpha}(E; \vec{x},\vec{x}')$ satisfies the equation 
\be
i\frac{d}{dx}
\hat{F}=\left[\frac{M M^\dag}{2|E|}+V(\vec{x})\right]\!\hat{F}\,, 
\qquad {\rm where} \qquad \frac{d}{dx}\equiv\hat{\vec{r}}\cdot
\!\boldsymbol{\nabla}\,.
\label{eq:schroed0}
\ee
Here $\hat{\vec{r}}$ is the unit vector in the direction of the neutrino 
propagation: $\hat{\vec{r}}\equiv (\vec{x}-\vec{x}')/|\vec{x}-\vec{x}'|$, 
and $\frac{d}{dx}$ as the directional derivative along $\vec{\hat{r}}$.
The effective potential $V(\vec{x})$ is related to the components 
of the neutrino potential in matter $V^\mu(\vec{x})$ through  
\be
V\equiv V^0-\boldsymbol{v}_\nu\!\cdot\!\vec{V}\simeq V^0-V^3\,,
\label{eq:ident}
\ee
where $\boldsymbol{v}_\nu$ is the neutrino velocity vector and $V^3$ is the 
component of $\vec{V}$ in the direction of neutrino propagation. The 
potential for antineutrinos is obtained from that for neutrinos by flipping 
the sign of the latter (except in CP-symmetric or nearly CP-symmetric media, 
see Appendix B). Taking into account that $M M^\dag=U M_d^2 U^\dag$ (which is 
valid in both the Dirac and Majorana neutrino cases), and that the neutrino 
potentials that enter in eqs.~(\ref{eq:schroed0}) 
and~(\ref{eq:Sch2}) are defined in the same way, 
we find that these equations coincide. 
Thus, the quantity $\hat{F}_{\beta\alpha}$ satisfies the same equation as 
the amplitude of $\nu_\alpha\to \nu_\beta$ oscillations in the standard 
approach to neutrino oscillations in matter.  
We summarize the expressions for the potential 
$V(x)$ for neutrino propagation in various media in Appendix B.

Let us now return to eq.~(\ref{eq:amp8}). We have defined $\vec{p}_*$ and 
$\vec{p}_*'$ as the momenta, small neighbourhoods of which give the main 
contributions to the integrals over $\vec{p}$ and $\vec{p}'$ in~(\ref{eq:S2}). 
How can one find these momenta? In the case of vacuum neutrino oscillations, 
the Grimus-Stockinger theorem tells us that due to a fast oscillating phase 
factor in the integrand of the Fourier-integral representation of the 
coordinate-space neutrino propagator, the neutrino is forced to be on its mass 
shell. Hence, $\Phi_{P}(E, \vec{p})$ and $\Phi_{D}(E,\vec{p}')$ should 
also be the production and detection amplitudes for on-shell neutrinos. This 
is a simple consequence of the fact that particles propagating macroscopic 
distances are essentially on their mass shells. One can therefore expect that 
for neutrino oscillations in non-uniform matter, in the case when neutrinos 
propagate macroscopic distances the production and detection amplitudes 
$\Phi_{P}(E, \vec{p})$ and $\Phi_{D}(E,\vec{p}')$ should also be taken on the 
``in-matter mass shells'' corresponding to the neutrino production and 
detection points, respectively. Here by the ``in-matter mass shell'' we mean 
that the neutrino energy and momentum at a fixed point with coordinate 
$\vec{x}$ should satisfy a dispersion relation that follows from the neutrino 
evolution equation (\ref{eq:schroed0}) with the effective 
potential $V(\vec{x})$. 
Thus, we expect that the momenta $\vec{p}_*$ and $\vec{p}_*'$ should satisfy 
the in-matter dispersion relations with the 
potentials $V(\vec{x}_P)$ and $V(\vec{x}_D)$, respectively. 
A direct proof of this statement will be given in Appendix C. 

Let us now show that the in-matter dispersion relations are well defined 
for neutrino production and detection. Indeed, the sizes of the localization 
regions of the neutrino production and detection processes, $\sigma_{xP}$ and 
$\sigma_{xD}$,  are small in comparison with the typical 
distances over which the neutrino potential in matter $V(\vec{x})$ varies 
significantly. Therefore, to a very good accuracy one can consider the 
production and detection processes as occurring at constant densities given 
by the matter densities at, respectively, neutrino production and detection 
points $\vec{x}_P$ and $\vec{x}_D$.

Note that the transverse components of the neutrino momentum 
are extremely small, $|\vec{p}_\perp|/p\sim {\rm max}\{\sigma_{xP},
\sigma_{xD}\}/L\lesssim 10^{-13}$ (see footnote \ref{foot:f1}), and so they 
can be safely neglected. As shown in Appendix C, the longitudinal components 
of the characteristic momenta $\vec{p}_*$ and $\vec{p}_*'$ satisfy 
eqs.~(\ref{eq:eq2c}) and (\ref{eq:eq2d}). Since these are matrix equations, 
it is convenient to go to the basis where the matrix $H$ is diagonal. 
For this purpose, we introduce the neutrino mixing matrix in matter 
according to
\be
|\nu_\alpha\rangle=\sum_K 
\tilde{U}_{\alpha K}^*(\vec{x})|\nu_K(\vec{x})\rangle\,,
\vspace*{-2.0mm}
\label{eq:mix2}
\ee
where 
$\tilde{U}_{\alpha K}(\vec{x})$ is the unitary matrix that diagonalizes the 
matrix $H(\vec{x})=MM^\dag/2|E|+V(\vec{x})$:
\be
H(\vec{x})=\tilde{U}(\vec{x}) {\cal H}(\vec{x})\tilde{U}(\vec{x})^\dag\,, 
\qquad {\cal H}(\vec{x})={\rm diag}\{{\cal H}_1(\vec{x}), {\cal H}_2
(\vec{x}),\dots\}\,.
\label{eq:diag}
\ee
The states $|\nu_K(\vec{x})\rangle$ are thus the local eigenstates of 
$H(\vec{x})$, which are called local matter eigenstates. Eq.~(\ref{eq:mix2}) 
relates the neutrino flavour eigenstate basis to the basis of the local 
matter eigenstates $|\nu_K(\vec{x})\rangle$, just like eq.~(\ref{eq:mix1}) 
relates it to the  mass eigenstate basis. 
In the limit of vanishing matter density the mixing matrix in matter 
$\tilde{U}(\vec{x})$ goes to the vacuum mixing matrix $U$ and the matter 
eigenstates go to the mass eigenstates. Note that eq.~(\ref{eq:mix2}) 
merely describes 
a basis transformation in eq.~(\ref{eq:Sch2}); 
it does not necessarily define the matter-eigenstate content of the produced 
and detected neutrino flavour states. It only does so when the neutrino 
production and detection coherence conditions are satisfied. We will discuss 
this point in more detail in section \ref{sec:evoleq}.

We also introduce the neutrino propagator in the matter eigenstate basis 
$\hat{{\cal S}}_{K' K}(E;\vec{x}', \vec{x})$. According to eq.~(\ref{eq:mix2}), 
it is related to the flavour-basis propagator $\hat{S}_{\beta\alpha}
(E;\vec{x}', \vec{x})$ through 
\be
\hat{S}_{\beta\alpha}(E;\vec{x}', \vec{x})=
\sum_{K,K'}\tilde{U}_{\beta K'}(\vec{x}')\tilde{U}_{\alpha K}^*(\vec{x})
\hat{{\cal S}}_{K' K}(E;\vec{x}', \vec{x})
=[\tilde{U}(\vec{x}')\hat{{\cal S}}(E;\vec{x}', \vec{x})
\tilde{U}^\dag(\vec{x})]_{\beta\alpha}\,.
\vspace*{-1mm}
\label{eq:mattS}
\ee
{}From (\ref{eq:propF1}) it follows that 
there is a similar relation between $\hat{F}_{\beta\alpha}$ and the 
corresponding matter-eigenstate quantity $\hat{{\cal F}}_{K' K}$: 
\be
\hat{F}_{\beta\alpha}(E;\vec{x}', \vec{x})=
\sum_{K,K'}\tilde{U}_{\beta K'}(\vec{x}')\tilde{U}_{\alpha K}^*(\vec{x})
\hat{{\cal F}}_{K' K}(E;\vec{x}', \vec{x})
=[\tilde{U}(\vec{x}')\hat{{\cal F}}(E;\vec{x}', \vec{x})
\tilde{U}^\dag(\vec{x})]_{\beta\alpha}\,.
\vspace*{-1mm}
\label{eq:mattF}
\ee
Going in eqs.~(\ref{eq:eq2c}) and (\ref{eq:eq2d}) to the matter-eigenstate 
basis allows one to immediately solve them with respect to the momenta 
${\rm p}_*$ and ${\rm p}'_*$, which are the longitudinal components 
of $\vec{p}_*$ and $\vec{p}'_*$. The results are given in eqs.~(\ref{eq:eq3a}) 
and (\ref{eq:eq3b}). Since the transverse components of $\vec{p}_*$ and 
$\vec{p}'_*$ essentially vanish, we conclude that the in-matter neutrino 
dispersion relations fully define the momenta 
that give main contributions to the integrals over $\vec{p}$ and $\vec{p}'$ 
in (\ref{eq:S2}). 
Indeed, eqs.~(\ref{eq:eq3a})-(\ref{eq:eq3c}) imply $\vec{p}_*=\vec{p}_K$ and 
$\vec{p}_*'=\vec{p}'_{K'}$, where $\vec{p}_K$ and $\vec{p}'_{K'}$ are the 
momenta of neutrino matter eigenstates at the production and detection points 
respectively.  
With this identification of $\vec{p}_*$ and $\vec{p}_*'$, 
eq.~(\ref{eq:amp8}) can be rewritten as 
\be
{\cal A}_{\beta\alpha}=
\sum_{K,K'}\tilde{U}_{\beta K'}(\vec{x}_D)\tilde{U}_{\alpha K}^*(\vec{x}_P)
\int\frac{dE}{2\pi}\Phi_{D}(E,\vec{p}'_{K'})\,\Phi_{P}(E, \vec{p}_K)\,
e^{-iE(t_D-t_P)}\hat{{\cal S}}_{K' K}(E;\vec{x}_D, \vec{x}_P)\,. 
\label{eq:amp8a}
\ee
This is the expression that we will be using in the following.

\section{\label{sec:prob}Total rate of the process and the oscillation 
probability}

In the previous section we calculated the amplitude of the overall 
neutrino production, propagation and detection process. Our next goal is 
to calculate the probability of this process and then extract from it  
the oscillation probability. 

Let us first recall how the oscillation probability is determined from 
experimental data. Assume that, in an experiment, neutrinos of flavour 
$\alpha$ are emitted by a source, with the neutrino production rate and 
energy spectrum being $\Gamma_\alpha^{\rm prod}$ and 
$d\Gamma_\alpha^{\rm prod}(E)/dE$. Let a detector sensitive to $\nu_\beta$ 
be situated at a distance $L$ from the source, and the detection cross section 
be $\sigma_\beta(E)$. The rate of the detection process is then  
\be
\Gamma^{\rm det}_{\alpha\beta}=\int dE \, j_\beta(E)\sigma_\beta(E)\,.
\label{eq:GammaD}
\ee
Here $j_\beta(E)$ is the flux of $\nu_\beta$ at the detector site, which 
is given by  
\be
j_\beta(E)=\frac{1}{4\pi L^2}\frac{d\Gamma_\alpha^{\rm prod}(E)}{dE}
P_{\alpha\beta}(E,\vec{x}_{D},\vec{x}_{P})\,, 
\label{eq:jbeta}
\ee
where $P_{\alpha\beta}(E,\vec{x}_{D},\vec{x}_{P})$ is the neutrino oscillation 
probability, and we assumed for simplicity that neutrino emission is 
isotropic. Substituting ({\ref{eq:jbeta}) into (\ref{eq:GammaD}) yields the 
rate of the overall production-propagation-detection process: 
\be
\Gamma^{\rm tot}_{\alpha\beta}\,\equiv\int dE\,
\frac{d\Gamma^{\rm tot}_{\alpha\beta}(E)}{dE}\,=\,\frac{1}{4\pi L^2}
\int dE\,\frac{d\Gamma_\alpha^{\rm prod}(E)}{dE}\,P_{\alpha\beta}
(E,\vec{x}_{D},\vec{x}_{P})\,\sigma_\beta(E)\,.
\label{eq:GammaTot}
\ee
If the spectral density of the overall process rate 
$d\Gamma^{\rm tot}_{\alpha\beta}(E)/dE$ is experimentally measured, one 
can find the oscillation probability by dividing this spectral density  
by the production rate, detection cross section and the geometric factor 
$1/4\pi L^2$: 
\be
P_{\alpha\beta}(E,\vec{x}_{D},\vec{x}_{P})\,=\,
\frac{d\Gamma^{\rm tot}_{\alpha\beta}(E)/dE}{\frac{1}{4\pi L^2}\,
[d\Gamma_\alpha^{\rm prod}(E)/dE]\,\sigma_\beta(E)}\,.
\label{eq:P2}
\ee
Notice that an important ingredient of this argument is the assumption that, 
for a fixed neutrino energy, the overall rate of the process factorizes into 
the production rate, oscillation probability and detection cross section. 
If such a factorization turns out to be impossible, the very notion of the 
oscillation probability loses its sense, and one has to deal instead with the 
probability of the overall process. 

Now, we shall calculate the rate of the overall process in our QFT-based 
approach and try to present it in a form similar to (\ref{eq:GammaTot}), which 
would allow us to find the oscillation probability. In doing so, we shall be 
closely following the treatment of the vacuum oscillations case in section 
5.2 of ref.~\cite{AK}, to which we refer the reader for details.  

Let us first calculate the spectral density of the production rate 
$d\Gamma_\alpha^{\rm prod}(E)/dE$ and the detection cross section 
$\sigma_\beta(E)$. 
To simplify the calculation, we will be assuming that the neutrino emission 
and absorption processes are isotropic (relaxing this assumption would 
complicate the analysis but would  not change the final result for the
probability of neutrino oscillations). This means that we can average the 
production and detection amplitudes over the direction of the incoming 
particles $P_i$ and $D_i$, which amounts to averaging over the directions of 
$\vec{L}=\vec{x}_D-\vec{x}_P$. One can therefore define $\Phi_P(E,{\rm p}_K)=
\int\frac{d\Omega_{\vec{L}}}{4 \pi}\Phi_P(E,\vec{p}_{K})$, 
$\Phi_D(E,{\rm p}'_{K'})=\int\frac{d\Omega_{\vec{L}}}{4 \pi}
\Phi_D(E,\vec{p}'_{K'})$. Applying the standard QFT rules, one then finds 
for the neutrino production and detection probabilities  
\begin{align}
 P^{\rm prod}_\alpha =\sum_K |\tilde{U}_{\alpha K}|^2 
& \int\frac{d^3 p_K}{(2\pi)^3}\,\big|\Phi_{P}(E,{\rm p}_K)\big|^2
=\sum_K |\tilde{U}_{\alpha K}|^2 \frac{1}{2\pi^2}\int 
dE\,\big|\Phi_{P}(E,{\rm p}_{K})\big|^2 E {\rm p}_K\,, 
\label{eq:prodrate}
\\[0.3em]
& P_\beta^{\rm det}(E)=\sum_{K'} |\tilde{U}_{\beta K'}|^2 
|\Phi_{D}(E,{\rm p}'_{K'})|^2 
\frac{1}{V_{N}}\,,
\label{eq:detProb}
\end{align}
where $V_{N}$ is the normalization volume, and ${\rm p}_K$, ${\rm p}'_{K'}$ 
are the energy-dependent momenta of neutrino matter eigenstates for 
$V(\vec{x})=V(\vec{x}_P)$ and $V(\vec{x})=V(\vec{x}_D)$, respectively, which 
are given by eqs.~(\ref{eq:eq3a}) and (\ref{eq:eq3b}).  
In eqs.~(\ref{eq:prodrate}), (\ref{eq:detProb}) and in the following we use 
the shorthand notation 
\be
\tilde{U}_{\alpha K}\equiv \tilde{U}_{\alpha K}(\vec{x}_P)\,,\qquad  
\tilde{U}_{\beta K'}\equiv \tilde{U}_{\beta K'}(\vec{x}_D)\,,
\label{eq:shorthand}
\ee
i.e.\ $\alpha$ and $K, M,...$ will always refer to, respectively, the flavour 
index of the produced neutrino state and the indices of its matter-eigenstate 
components, whereas $\beta$ and $K', M',...$ will similarly refer to the 
flavour and mass-eigenstate components of the detected state. 

{}From eqs.~(\ref{eq:prodrate}) and (\ref{eq:detProb}) one can find the 
spectral density of the produced neutrino flux and the detection cross 
section \cite{AK}:
\begin{align}
& \frac{d\Gamma_\alpha^{\rm prod}(E)}{dE}=\frac{N_P}{T_0}\sum_K 
|\tilde{U}_{\alpha K}|^2 \frac{1}{2\pi^2} \big|\Phi_{P}(E,{\rm p}_{K})\big|^2 
E {\rm p}_K\,, 
\label{eq:ratedens}
\\[0.2em]
& \sigma_\beta(E)= \frac{N_D}{T_0} \sum_{K'} |\tilde{U}_{\beta K'}|^2
|\Phi_{D}(E,{\rm p}'_{K'})|^2 \frac{E}{{\rm p}'_{K'}}\,.
\label{eq:sigmabeta}
\end{align}
Here $N_P/T_0$ and $N_D/T_0$ are flux-dependent normalization constants 
\cite{AK}, which will drop out of the final result for the oscillation 
probability.  

Next, we need the rate of the overall neutrino 
production-propagation-detection process, which can be found by integrating 
the squared modulus of the amplitude of the process over the production and 
detection times $t_P$ and $t_D$. The time integrals can be reduced to the 
integrals over $(t_P+t_D)/2$ and $T\equiv t_D-t_P$. The first integration is 
trivial, whereas the second one leads to 
\be
\Gamma_{\alpha\beta}^{\rm tot}=\frac{N_P N_D}{T_0^2}\,
\int dT|{\cal A}_{\beta\alpha}(T,\vec{x}_D,\vec{x}_P)|^2\,.
\ee
Substituting here the expression for the amplitude ${\cal A}_{\beta\alpha}
(T,\vec{x}_D,\vec{x}_P)$ from (\ref{eq:amp8a}), we find 
\begin{align}
\Gamma_{\alpha\beta}^{\rm tot}=\frac{N_P N_D}{T_0^2}\frac{1}{(4\pi)^2 L^{2}}
\int\!\frac{dE}{2\pi}(2 E)^2 \!\!
\sum_{K,K',M,M'}\!\!\tilde{U}_{\alpha K}^* \tilde{U}_{\beta K'}^{}
\tilde{U}_{\alpha M}^{} \tilde{U}_{\beta M'}^* 
\Phi_{D}(E,{\rm p}'_{K'}) \Phi_{P}(E,{\rm p}_{K})\,
\nonumber \\[0.2em]
\times\Phi^*_{D}(E,{\rm p}'_{M'}) \Phi^*_{P}(E,{\rm p}_{M})\hat{\cal F}_{K'K}
(E;\vec{x}_D, \vec{x}_P)
\hat{\cal F}^{*}_{M'M}(E;\vec{x}_D, \vec{x}_P)\,.
\label{eq:totrate}
\end{align}
The quantity $\hat{\cal F}_{K'K}(E;\vec{x}_D,\vec{x}_P)$ introduced here 
is related to $\hat{\cal S}_{K'K}(E;\vec{x}_D,\vec{x}_P)$ in the same way as 
$\hat{F}_{K'K}(E;\vec{x}_D,\vec{x}_P)$ is related 
to $\hat{S}_{K'K}(E;\vec{x}_D,\vec{x}_P)$ (see eq.~(\ref{eq:propF1})). 
The spectral density $d\Gamma_{\alpha\beta}^{\rm tot}(E)/dE$ is obtained from 
the right hand side of 
eq.~(\ref{eq:totrate}) by removing the integration over $E$. 

By comparing eq.~(\ref{eq:totrate}) with eqs.~(\ref{eq:ratedens}) and 
(\ref{eq:sigmabeta}) it can be seen that the factorization of the rate of 
the overall process into the production rate, propagation (oscillation) 
probability and detection cross section as in eq.~(\ref{eq:GammaTot}) is 
only possible if the production and detection amplitudes $\Phi_P$, $\Phi_D$ 
can be pulled out of the sum in (\ref{eq:totrate}). This, in turn, is allowed 
only if the corresponding momenta of the matter eigenstates 
\vspace*{-2mm}satisfy %
\footnote{While conditions (\ref{eq:Condit1}) and (\ref{eq:Condit2}) ensure 
the production and detection coherence, they say nothing about another possible 
source of decoherence -- separation of neutrino wave packets at long 
enough distances $L>L_{\rm coh}$ due to the difference of the group 
velocities of different neutrino mass eigenstates. This is related to
the fact that a fixed neutrino energy corresponds to the stationary 
situation, when the coherence length $L_{\rm coh}\to \infty$. The finite 
coherence length is recovered upon the integration over energy in 
eq.~(\ref{eq:totrate}) \cite{Beuthe1}.}
\begin{align}
&|{\rm p}_K-{\rm p}_M|\ll\sigma_{pP}\,,
\label{eq:Condit1}
\\
&|{\rm p}'_{K'}-{\rm p}'_{M'}|\ll\sigma_{pD}\,.
\label{eq:Condit2}
\end{align}
Indeed, under these conditions the factors $\Phi_{P}(E,{\rm p}_K)$ and 
$\Phi_{D}(E,{\rm p}'_{K'})$ are essentially independent of the indices $K$ 
and $K'$; one can therefore replace them, 
respectively, by the quantities $\Phi_{P}(E,{\rm p})$ and 
$\Phi_{D}(E,{\rm p}')$ calculated at the mean momenta ${\rm p}$ and ${\rm p}'$  
and pull them out of the sum. From eq.~(\ref{eq:totrate}) we then find 
\begin{align}
\frac{d\Gamma_{\alpha\beta}^{\rm tot}(E)}{dE}=\frac{N_P 
N_D}{T_0^2}& \frac{1}{(4\pi)^2 L^{2}} 
|\Phi_{P}(E,{\rm p})|^2\,|\Phi_{D}(E,{\rm p}')|^2 
\nonumber \\[0.3em]
\times & \sum_{K,K',M,M'}\!\!\tilde{U}_{\alpha K}^* \tilde{U}_{\beta K'}^{}
\tilde{U}_{\alpha M}^{} \tilde{U}_{\beta M'}^* \hat{\cal F}_{K'K}(E;\vec{x}_D, 
\vec{x}_P)\hat{\cal F}^{*}_{M'M}(E;\vec{x}_D, \vec{x}_P)\,.
\label{eq:totrate1}
\end{align}
Likewise, under conditions (\ref{eq:Condit1}) and (\ref{eq:Condit2})
one can replace $\Phi_{P}$ and $\Phi_{D}$ as well as the factors 
${\rm p}_K$ and $1/{\rm p}_{K'}$ in 
eqs.~(\ref{eq:ratedens}) and (\ref{eq:sigmabeta}) by the corresponding 
quantities taken at the average values of the relevant momenta. They can 
then be pulled out of the sums, which yields
\begin{align}
& \frac{d\Gamma_\alpha^{\rm prod}(E)}{dE}=\frac{N_P}{T_0} 
\frac{1}{2\pi^2} \big|\Phi_{P}(E,{\rm p})\big|^2 
E {\rm p}\,, 
\label{eq:ratedens1}
\\[0.4em]
& \sigma_\beta(E)= \frac{N_D}{T_0} 
|\Phi_{D}(E,{\rm p}')|^2 \frac{E}{{\rm p}'}\,.
\label{eq:sigmabeta1}
\end{align}
Here we have used unitarity of the leptonic mixing matrix in matter 
$\tilde{U}$. Substituting these expressions, together with 
$d\Gamma_{\alpha\beta}^{\rm tot}(E)/dE$ from eq.~(\ref{eq:totrate1}),  
into (\ref{eq:P2}), we arrive at
\begin{align}
P_{\alpha\beta}(E, \vec{x}_D, \vec{x}_P)=
\displaystyle{\sum_{{K,K',M,M'}}}\tilde{U}_{\alpha K}^* 
\tilde{U}_{\beta K'}\tilde{U}_{\alpha M} \tilde{U}_{\beta M'}^* 
&\hat{\cal F}_{K'K}(E;\vec{x}_D, \vec{x}_P)\hat{\cal F}^*_{M'M}(E; \vec{x}_D, 
\vec{x}_P)
\nonumber \\
&=|\hat{F}_{\beta\alpha}(E;\vec{x}_D, \vec{x}_P)|^2\,, 
\label{eq:probSimpl}
\end{align}
where the flavour-basis function $\hat{F}_{\beta\alpha}(E;\vec{x}_D, 
\vec{x}_P)$ obeys eq.~(\ref{eq:schroed0}) with the boundary condition 
(\ref{eq:boundary}). Here the factors $|\Phi_P(E,{\rm p})|^2 
|\Phi_D(E,{\rm p}')|^2$ in the numerator and denominator have canceled out, 
leaving us with the oscillation probability that is independent of the 
neutrino production and detection processes. In deriving 
eq.~(\ref{eq:probSimpl}) we have also canceled ${\rm p}$ and ${\rm p}'^{-1}$ 
in the product $(d\Gamma_\alpha^{\rm prod}(E)/dE)\times\sigma_\beta(E)$ 
in the denominator. This is justified because the mean neutrino momenta at 
production and detection coincide to a very good accuracy under the conditions 
$\Delta m^2/(2E)\ll E$, $|V|\ll |E|$, which we assume to be satisfied 
throughout this paper. 

Thus, we have found that under conditions (\ref{eq:Condit1})
and (\ref{eq:Condit2}) the oscillation probability can be sensibly defined 
and can be extracted from the rate of the overall neutrino 
production-propagation-detection process. Since the matrix $\hat{F}$ is 
unitary,%
\footnote{This follows from the fact that $\hat{F}$ satisfies the 
Schr\"odinger-like equation (\ref{eq:schroed0}) with the Hermitian 
effective Hamiltonian, supplemented the boundary condition 
(\ref{eq:boundary}).}   
the resulting oscillation probability~(\ref{eq:probSimpl}) obeys  
the unitarity conditions $\sum_\beta P_{\alpha\beta}(E, \vec{x}_D, \vec{x}_P)=
\sum_\alpha P_{\alpha\beta}(E, \vec{x}_D, \vec{x}_P)=1$, i.e.\ is properly 
normalized. If conditions  (\ref{eq:Condit1}) and (\ref{eq:Condit2}) are 
not fulfilled, the oscillation probability cannot be defined, and flavour 
transitions 
should instead be described by the rate of the overall neutrino 
production-propagation-detection process~(\ref{eq:totrate}).

Eqs.~(\ref{eq:Condit1}) and (\ref{eq:Condit2}) are actually the 
conditions of coherent neutrino production and detection: their fulfilment 
ensures that the production and detection processes cannot distinguish between 
different neutrino matter eigenstates, so that these eigenstates are produced 
and detected coherently. If these conditions are violated, i.e.\ if either 
$|{\rm p}_K-{\rm p}_M|\gtrsim \sigma_{pP}$ or $|{\rm p}'_{K'}-{\rm p}'_{M'}|
\gtrsim \sigma_{pD}$, the differences of momenta of different matter 
eigenstates will exceed the momentum widths of the corresponding momentum 
distribution amplitudes, $\Phi_{P}$ or $\Phi_{D}$. In that case the overlap 
of the amplitudes corresponding to different matter eigenstates will be 
suppressed, leading to a quenching of the interference terms in 
expression~(\ref{eq:totrate}) for the probability of the overall process.  
Note that the momentum uncertainties due to the localization of the 
neutrino production and detection processes, $\sigma_{pP}$ and $\sigma_{pD}$, 
are usually much smaller than the neutrino momentum itself; therefore, 
conditions in eqs.~(\ref{eq:Condit1}) and (\ref{eq:Condit2}) are much stronger 
than the conditions $|{\rm p}_K-{\rm p}_M|\ll {\rm p}_K, {\rm p}_M$, 
$|{\rm p}'_{K'}-{\rm p}'_{M'}|\ll {\rm p}'_{K'}, {\rm p}'_{M'}$, which follow 
automatically from $\Delta m^2/(2E)\ll E$, $|V|\ll |E|$.

\section{
\label{sec:evoleq}The amplitude of the overall process, the oscillation 
amplitude and their evolution equations}

We have demonstrated in the previous section that in the case when neutrinos 
are ultra-relativistic, the matter-induced neutrino potential satisfies 
$|V(\vec{x})|\ll |E|$, and in addition the conditions of coherent neutrino 
production and detection (\ref{eq:Condit1}) and (\ref{eq:Condit2}) are 
fulfilled, the oscillation probability can be sensibly defined and can be 
extracted from the rate of the overall neutrino 
production-propagation-detection process. The resulting expression for the 
oscillation probability in eq.~(\ref{eq:probSimpl}) is simply given by the 
squared modulus of $\hat{F}_{\beta\alpha}$, which therefore can be interpreted 
as the oscillation amplitude. 
As we have already discussed, $\hat{F}_{\beta\alpha}$
satisfies the evolution equation (\ref{eq:schroed0}) 
(which coincides with eq.~(\ref{eq:Sch2})), supplemented by the boundary 
condition (\ref{eq:boundary}). Thus, in the 
case when the coherence conditions for neutrino production and detection 
are satisfied, the standard approach to neutrino oscillations in matter 
based on the Schr\"odinger-like evolution equation (\ref{eq:Sch2}) is 
justified. 
 
Let us now discuss the amplitude of the overall neutrino 
production-propagation-detection process. Does it satisfy an evolution 
equation similar to~(\ref{eq:Sch2})? 
Consider first the case of vacuum 
neutrino oscillations. The neutrino production and detection coherence 
conditions now read 
\be
|{\rm p}_j-{\rm p}_k|\simeq |\Delta m_{jk}^2/(2E)|\ll 
\sigma_{pP}\,,\sigma_{pD}\,.
\label{eq:Condit3}
\ee
Here $p_j=({E^2-m_j^2})^{1/2}\approx E-m_j^2/(2E)$ is the momentum of the 
$j$th neutrino mass eigenstate of energy $E$.
If these conditions are satisfied, the oscillation amplitude can be defined, 
and it coincides with the standard amplitude of neutrino oscillations in 
vacuum: 
\be
[{\cal A}_{\rm vac}^{\rm osc}(E,x)]_{\beta\alpha}=
\sum\nolimits_{j}{U}_{\alpha j}^* {U}^{}_{\beta j} e^{-i\frac{\Delta 
m_{jk}^2}{2E}x}\,.
\label{eq:ampl4c}
\ee
The amplitude (\ref{eq:ampl4c}) satisfies the Schr\"odinger-like equation 
\be
i\frac{d}{dx}{\cal A}^{\rm osc}_{\rm vac}(E,x)=\Big[U\frac{\Delta 
m^2}{2E}U^{\dag}\Big] {\cal A}^{\rm osc}_{\rm vac}(E,x)\,
\label{eq:schroedlike}
\ee
with the boundary condition $[A^{\rm osc}_{\rm vac}(E,0)]_{\beta\alpha}=
\delta_{\beta\alpha}$.  

Let us now examine the probability of the overall neutrino 
production-propagation-detection process in vacuum, without assuming 
anything about coherence of neutrino production and detection. This 
probability can be written as \cite{AK} 
\be
\Gamma_{\alpha\beta}^{\rm tot}(x)=\frac{N_P N_D}{T_0^2}
\int \frac{dE}{2\pi}(2E)^2 |{\cal A}^{\rm tot}_{\rm 
vac}(E,x)_{\beta\alpha}|^2\,,
\label{eq:totratevac}
\ee
where the quantity 
\be
{\cal A}^{\rm tot}_{\rm vac}(E,x)_{\beta\alpha}\equiv
\sum_j U_{\alpha j}^{*} U_{\beta j}^{}
\Phi_P(E,{\rm p}_j) \Phi_D(E,{\rm p}_j) e^{i(p_j-p_1)x}=
\big\{U [\Phi_P \Phi_D e^{i\Delta p\cdot  
x}]U^\dag\big\}_{\beta\alpha}\,
\label{eq:ampvac}
\ee
can be considered as the amplitude of the overall process. It is 
eqs.~(\ref{eq:totratevac}) and (\ref{eq:ampvac}) that have to be used to 
describe neutrino flavour transitions in vacuum in the case when the coherence 
condition in eq.~(\ref{eq:Condit3}) are violated. From eq.~({\ref{eq:ampvac}) 
it is easy to find that the amplitude of the overall process in vacuum 
${\cal A}^{\rm tot}_{\rm vac}(E,x)$ satisfies the same evolution equation as 
the oscillation amplitude. Indeed, differentiating (\ref{eq:ampvac}) we obtain 
$i(d/dx){\cal A}^{\rm tot}_{\rm vac}=U[\Phi_P \Phi_D e^{i\Delta p x}
(-\Delta p)]U^\dag=U(-\Delta p)U^\dag U [\Phi_P \Phi_D e^{i\Delta p x}]
U^\dag=U(-\Delta p)U^\dag {\cal A}^{\rm tot}_{\rm vac}$, which coincides with 
(\ref{eq:schroedlike}). Crucial to this derivation was the point that all the 
factors in the square brackets are diagonal and therefore commute with 
each other. 

Although ${\cal A}^{\rm osc}_{\rm vac}(E,x)$ and ${\cal A}^{\rm tot}_{\rm 
vac}(E,x)$ satisfy the same evolution equations, the boundary conditions 
that they obey are different. As was mentioned above, for the oscillation 
amplitude it is the standard condition 
$[A^{\rm osc}_{\rm vac}(E,0)]_{\beta\alpha}=\delta_{\beta\alpha}$; at the 
same time, for the overall amplitude the boundary condition is 
$[A_{\rm osc}^{\rm tot}(E,0)]_{\beta\alpha}=
\{U\Phi_P\Phi_DU^\dag\}_{\beta\alpha}$, as can be immediately seen from 
eq.~(\ref{eq:ampvac}). Obviously, the solution of one and the same 
eq.~(\ref{eq:schroedlike}) with two different boundary conditions are 
different. 

Now let us return to neutrino oscillations in matter. The rate of the overall 
process (\ref{eq:totrate}) can be cast in the same form as in 
eq.~(\ref{eq:totratevac}), but with the vacuum amplitude
${\cal A}^{\rm tot}_{\rm vac}(E,x)_{\beta\alpha}$ replaced by 
\begin{align}
{\cal A}^{\rm tot}_{\beta\alpha}(E, \vec{x},\vec{x}_0)\equiv
\sum_{K,K'} \tilde{U}_{\alpha K}^*(\vec{x}_0) \tilde{U}_{\beta 
K'}^{}(\vec{x})
\Phi_P(E,{\rm p}_K) \Phi_D(E,{\rm p}'_{K'}) 
\hat{\cal F}_{K' K}(E; \vec{x},\vec{x}_0)
\nonumber \\
= \big\{\tilde{U}(\vec{x}) \Phi_D \hat{\cal F}\Phi_P
\tilde{U}^\dag(\vec{x}_0)\big\}_{\beta\alpha}\,.
\label{eq:amptotmatt}
\end{align}
This expression has a simple physical interpretation: the factor 
$\tilde{U}^\dag(\vec{x}_0)$ projects the initial flavour-eigenstate neutrino 
$\nu_\alpha$ onto the matter eigenstate basis, $\Phi_P(E,{\rm p}_K)$ are the 
amplitudes of production at the point $\vec{x}_0$ of various matter 
eigenstates that compose $\nu_\alpha$, $\hat{\cal F}(E;\vec{x}, \vec{x}_0)$ 
describes the propagation of these matter eigenstates to the point $\vec{x}$ 
(including the transitions between them), $\Phi_D(E,{\rm p}'_{K'})$ are the 
detection amplitudes of neutrino matter eigenstates at the point $\vec{x}$, 
and finally $\tilde{U}(\vec{x})$ projects the amplitude back from 
the matter eigenstate basis to the flavour basis.

Consider the case when both the coherence conditions (\ref{eq:Condit1}) 
and (\ref{eq:Condit2}) are violated, so that the amplitudes $\Phi_P$ 
and $\Phi_D$ cannot be pulled out of the sum in (\ref{eq:amptotmatt}). 
Does the amplitude of the overall process ${\cal A}^{\rm tot}$ satisfy the 
same evolution equation as the quantity $\hat{F}$, as it is the case for 
neutrino oscillations in vacuum? By differentiating eq.~(\ref{eq:amptotmatt}) 
with respect to $x$,%
\footnote{Recall that $d/dx$ here is understood not as the derivative with 
respect to $|\vec{x}|$, but as a directional derivative along  
${\vec r}\equiv \vec{x}-\vec{x}_0$, see eq.~(\ref{eq:schroed0}).}
we immediately find that in general this is {\em not} the case. The 
reason for this is 
that, unlike in the case of vacuum oscillations, the matrix 
$\hat{\cal F}(E;\vec{x},\vec{x}_0)$ is not diagonal. Actually, for  
neutrinos moving in non-uniform matter the neutrino propagator is 
not diagonal in any basis. 
This comes about because the effective Hamiltonian $H(\vec{x})$ cannot be 
diagonalized by one and the same unitary transformation for all values of 
$\vec{x}$. The only exception is the special case of adiabatic neutrino 
evolution, when the 
propagator is diagonal in the matter eigenstate basis. In this case, by 
differentiating (\ref{eq:amptotmatt}) with respect to $x$ one can make sure 
that the oscillation amplitude satisfies the standard evolution equation 
(\ref{eq:Sch2}) (though with a non-standard boundary condition). The proof is 
very similar to the one in the case of vacuum neutrino oscillations 
and is given in Appendix~D. 

What happens in the situations when one of the coherence condition 
(\ref{eq:Condit1}), (\ref{eq:Condit2}) is satisfied, while the other 
is not? To answer this question, it will be convenient for us to rewrite 
eq.~(\ref{eq:amptotmatt}) in the form 
\be
{\cal A}^{\rm tot}_{\beta\alpha}(E,\vec{x},\vec{x}_0)=
\big\{
\tilde{U}(\vec{x}) \Phi_D \tilde{U}^\dag(\vec{x})\,
\hat{F}(E;\vec{x},\vec{x}_0)\,\tilde{U}(\vec{x}_0) \Phi_P 
\tilde{U}^\dag(\vec{x}_0)
\big\}_{\beta\alpha}\,,
\label{eq:amptotmatt2}
\ee
where we have used eq.~(\ref{eq:mattF}).
This expression admits a simple interpretation similar to that of  
eq.~(\ref{eq:amptotmatt}) (see below). 

Consider first the case when the detection coherence condition 
(\ref{eq:Condit2}) is satisfied, but the production coherence condition 
(\ref{eq:Condit1}) is violated. In this case one can replace 
the factors $\Phi_D(E, {\rm p}'_{K'})$ in eq.~(\ref{eq:amptotmatt2}) by 
the one taken at the mean momentum ${\rm p}'$, which yields 
\be
{\cal A}^{\rm tot}_{\beta\alpha}(E,\vec{x},\vec{x}_0)=
\Phi_D(E, {\rm p}')\big\{
\hat{F}(E;\vec{x},\vec{x}_0)\,\tilde{U}(\vec{x}_0) \Phi_P 
\tilde{U}^\dag(\vec{x}_0)\big\}_{\beta\alpha}\,.
\label{eq:amptotmatt3}
\ee
{}From the fact that only the first factor in the curly brackets here depends 
on $\vec{x}$, it immediately follows that in this case the amplitude 
${\cal A}^{\rm tot}(E,\vec{x},\vec{x}_0)$ satisfies the same 
equation as $\hat{F}(E;\vec{x},\vec{x}_0)$ does, i.e. eq.~(\ref{eq:schroed0}).
The boundary condition for the overall amplitude is, however, different:
from eq.~(\ref{eq:amptotmatt3}) we find 
\be
{\cal A}^{\rm tot}_{\beta\alpha}
(E,\vec{x},\vec{x}_0)|_{\vec{x}\to \vec{x}_0}=
\Phi_D(E, {\rm p}')\big\{\tilde{U}(\vec{x}_0) \Phi_P 
\tilde{U}^\dag(\vec{x}_0)\big\}_{\beta\alpha}\,.
\label{eq:boundary3}
\ee

Now let us consider the opposite case when the production coherence condition 
(\ref{eq:Condit1}) is satisfied, but the detection coherence condition 
(\ref{eq:Condit2}) is not. Then 
from (\ref{eq:amptotmatt2}) we find 
\be
{\cal A}^{\rm tot}_{\beta\alpha}(E,\vec{x},\vec{x}_0)=
\Phi_P(E, {\rm p})\big\{\tilde{U}(\vec{x}) \Phi_D \tilde{U}^\dag(\vec{x})\,
\hat{F}(E;\vec{x},\vec{x}_0)\big\}_{\beta\alpha}\,. 
\label{eq:amptotmatt4}
\ee
This expression contains, in addition to $\hat{F}(E;\vec{x},\vec{x}_0)$, two 
more $\vec{x}$-dependent factors, $\tilde{U}(\vec{x})$ and 
$\tilde{U}(\vec{x})^\dag$; it can be readily seen that the amplitude  
${\cal A}^{\rm tot}(E,\vec{x},\vec{x}_0)$ does not satisfy the 
same equation as $\hat{F}(E;\vec{x},\vec{x}_0)$ in this case. 

Thus, we found some disparity between the production and detection 
processes: if neutrino detection is coherent but the production process is 
incoherent, the amplitude of the overall process obeys the standard 
evolution equation (\ref{eq:Sch2}), while in the opposite situation it does 
not. The reason for this asymmetry is that we assume the neutrino production 
coordinate to be fixed and consider the evolution of the amplitude with 
the coordinate of the neutrino detection point. If the detection process is 
incoherent, the flavour-eigenstate detection amplitude $\tilde{U}(\vec{x})
\Phi_D \tilde{U}^\dag(\vec{x})$ is coordinate-dependent, and the 
$\vec{x}$-dependence of ${\cal A}^{\rm tot}(E,\vec{x},\vec{x}_0)$ is 
different from that of $\hat{F}(E;\vec{x},\vec{x}_0)$.  
Therefore the amplitude ${\cal A}^{\rm tot}(E,\vec{x},\vec{x}_0)$ does not 
satisfy eq.~(\ref{eq:Sch2}). 

How can one understand the above results in physical terms? Consider the 
matter eigenstate content of the initially produced neutrino state 
$\nu_\alpha$. The probability amplitude that the initial flavour state 
contains the matter eigenstate $\nu_K$ is given by $\Phi_P(E,{\rm p}_K)
\tilde{U}^*_{\alpha K}(\vec{x}_0)$. It differs 
from the naively expected factor $\tilde{U}^*_{\alpha K}(\vec{x}_0)$ that 
would follow from eq.~(\ref{eq:mix2}) by the presence of the $K$-dependent 
amplitude of $\nu_K$ production $\Phi_P(E,{\rm p}_K)$. In general, 
eq.~(\ref{eq:mix2}) should actually be considered as the definition of the 
matter eigenstate basis rather than a relation giving the matter-eigenstate 
composition of the flavour neutrino state, which is process dependent. This 
comes about because eq.~(\ref{eq:mix2}) describes the basis transformation in 
the evolution equation (\ref{eq:Sch2}) which ignores the coherence issues. 
If the production coherence condition (\ref{eq:Condit1}) is satisfied, all the 
amplitudes $\Phi_P(E,{\rm p}_K)$ can to a very good accuracy be replaced by a 
common factor $\Phi_P(E,{\rm p})$. In this case the relative weights of 
different matter eigenstates in $\nu_\alpha$ are given by  
$|\tilde{U}_{\alpha K}(\vec{x}_0|^2$, i.e.\  eq.~(\ref{eq:mix2}) does give   
the matter-eigenstate content of $\nu_\alpha$. 
If, on the contrary, condition (\ref{eq:Condit1}) is strongly violated, 
different matter eigenstates will be produced incoherently. Indeed, the 
squared modulus of the overall amplitude contains terms proportional to 
$\Phi_P(E,{\rm p}_K) \Phi_P^*(E,{\rm p}_M)$; 
for $K\ne M$ these are the interference terms. 
If $|{\rm p}_M-{\rm p}_K|$ is large compared to  
the momentum width $\sigma_{pP}$ of the amplitude $\Phi_P$, the quantities  
$\Phi_P(E,{\rm p}_K)$ and $\Phi_P^*(E,{\rm p}_M)$ will have little overlap. 
In this case the interference terms are strongly suppressed, which means 
that $\nu_K$ and $\nu_M$ are emitted incoherently.  

The initially produced neutrino state can then be evolved from $\vec{x}_0$ 
to $\vec{x}$ by $\hat{\cal F}(E;\vec{x},\vec{x}_0)$, as in 
eq.~(\ref{eq:amptotmatt}). 
Alternatively, one can project the initial state onto the flavour basis 
and evolve it with $\hat{F}(E;\vec{x},\vec{x}_0)$, as in 
eq.~(\ref{eq:amptotmatt2}). 
The evolved neutrino state is then absorbed in the detection process. Each of 
the matter eigenstate components $\nu_{K'}$ of the evolved state is detected 
with its own amplitude $\Phi_D(E,{\rm p}'_{K'})$, so that in general the 
matter-eigenstate composition of the detected state is not  
given by eq.~(\ref{eq:mix2}). Therefore the amplitude of the overall 
production-propagation-detection process does not satisfy the standard 
evolution equation (\ref{eq:Sch2}).
However, if all $\Phi_D(E,{\rm p}'_{K'})$ are to a good accuracy equal to 
each other (which is the case when the detection coherence condition 
(\ref{eq:Condit2}) is fulfilled), the detection efficiency is essentially 
the same for all matter eigenstates, so that the detected flavour state 
is indeed related to the matter eigenstates by eq.~(\ref{eq:mix2}). 
In this case the standard evolution equation 
(\ref{eq:Sch2}) applies. As follows from the above discussion, this holds  
irrespectively of  
whether or not the production coherence condition (\ref{eq:Condit1}) is 
obeyed. The latter just determines the initial state of neutrino evolution.  

In brief, if the matter-eigenstate composition of the evolving neutrino state 
is described by eq.~(\ref{eq:mix2}), the amplitude of the overall process 
evolves according to eq.~(\ref{eq:Sch2}). Otherwise, eq.~(\ref{eq:Sch2}) does 
not apply, the only exception being the case of adiabatic neutrino evolution.

There is an important remark that has to be added to the above discussion. 
We have found that in the case when neutrino detection 
is coherent while its production is not the amplitude of 
the overall process still satisfies evolution equation (\ref{eq:Sch2}).
However, even in this case the standard 
approach to neutrino oscillations in non-uniform matter has to be modified. 
This follows from the fact that the amplitude of the overall process does 
not factorize into the production, oscillation and 
detection amplitudes in this case; only the detection amplitude can be 
factored out. In such a situation one has to deal with the 
probability of the overall process, described by eq.~(\ref{eq:totrate}).
Alternatively, one can employ eq.~(\ref{eq:totratevac}) with the vacuum 
amplitude ${\cal A}^{\rm tot}_{\rm vac}(E,x)_{\beta\alpha}$ replaced by 
${\cal A}^{\rm tot}_{\beta\alpha}(E,x)$, where 
${\cal A}^{\rm tot}_{\beta\alpha}(E,x)$ satisfies eq.~(\ref{eq:Sch2}) 
with the boundary condition (\ref{eq:boundary3}).

\section{Discussion and summary\label{sec:disc}}

In this paper we have considered neutrino oscillations in non-uniform 
matter in the framework of QFT. We treated neutrino production, 
propagation and detection as a single process, described by the Feynman 
diagram of fig.~\ref{fig:feyn}, with neutrino in the intermediate state 
described by a propagator. 
We found that 
under certain conditions (which are satisfied in most cases of practical 
interest) the oscillation probability can be sensibly defined. 

We have demonstrated that when the conditions for the existence of the 
oscillation probability are fulfilled, this probability is given by 
eq.~(\ref{eq:probSimpl}). 
The oscillation amplitude in this case coincides with the function 
$\hat{F}_{\beta\alpha}(E; \vec{x},\vec{x}_0)$ that is simply related to the 
neutrino propagator in matter. 
This function satisfies the usual Schr\"odinger-like evolution equation 
(\ref{eq:schroed0}), provided that matter density is sufficiently smooth,  
so that condition (\ref{eq:condit1}) is satisfied. 
Hence, in this case the standard approach to neutrino oscillations in 
non-uniform matter 
is justified. We thus presented here a consistent derivation of the standard 
evolution equation, and found the conditions under which it is valid. Let us 
summarize here these condition once again: 

\begin{itemize}

\item[(i)]
Neutrinos are ultra-relativistic, 
so that $\frac{\Delta m^2}{2E}\ll E$.

\item[(ii)]
The effective matter-induced potential of neutrinos depends on the 
coordinate but does not vary with time, i.e.\ $V=V(\vec{x})$.

\item[(iii)]
The potential $V(\vec{x})$ is small compared to the mean neutrino 
energy: $|V(\vec{x})|\ll E$.

\item[(iv)] 
In the neutrino production and detection regions, 
matter density (and so the potential $V(\vec{x})$) is nearly 
constant. That is, $V(\vec{x})$ varies little over the distances of 
order of the sizes of the spatial localization regions of neutrino production 
and detection, $\sigma_{xP}$ and $\sigma_{xD}$. 
In other words, $|V'/V|\ll \min\{\sigma_{pP},\sigma_{pD}\}$, where 
$\sigma_{pP}\sim 1/\sigma_{xP}$ and $\sigma_{pD}\sim 1/\sigma_{xD}$ are the 
momentum uncertainties at neutrino production and detection, respectively.     

\item[(v)] Neutrino emission and detection are coherent, i.e.\ the 
conditions $|{\rm p}_K-{\rm p}_M|\ll\sigma_{pP}$ and 
$|{\rm p}'_{K'}-{\rm p}'_{M'}|\ll\sigma_{pD}$ are satisfied. 

\end{itemize}
In addition, when deriving eq.~(\ref{eq:schroed0}) we had to assume that the 
potential $V(\vec{x})$ varies little over the distances of order of the 
neutrino de~Broglie wavelength. 
However, since the momentum uncertainties at neutrino production and 
detection satisfy $\{\sigma_{pP},\sigma_{pD}\}\ll p$, this condition is 
superseded by the one in (iv), provided that 
the condition $|V'(\vec{x})/V(\vec{x})|\ll \min\{\sigma_{pP},\sigma_{pD}\}$ of 
point (iv) is fulfilled for all $\vec{x}$ along the neutrino trajectory and 
not only in the production and detection regions.

Conditions (i)-(v) ensure that the oscillation probability can be 
sensibly defined and can be extracted from the probability of the overall 
neutrino production-propagation-detection process. Condition (iii) allows to 
simplify significantly the equation for $\hat{F}(E;\vec{x}, \vec{x}_0)$ and 
reduce it to the form (\ref{eq:schroed1}). Condition (ii) simplifies 
the consideration, but in fact is not necessary. It is enough to assume 
that the neutrino wave packets are sufficiently short, 
so that the potential is nearly constant in space and time over the distances 
of order $\sigma_{x\nu}$ and times $\sim \sigma_{x\nu}/v_\nu$, where 
$\sigma_{x\nu}$ is the length of the neutrino wave packet and $v_\nu$ is its 
group velocity. Note that this assumption is related to condition (iv) because 
$\sigma_{x\nu}\lesssim 
\max\{\sigma_{pP}^{-1},\sigma_{pD}^{-1}\}$. Under the requirement that the 
potential $V(x)$ vary very little over the distances $\sim \sigma_{x\nu}$ 
and times $\sim \sigma_{x\nu}/v_\nu$ the neutrino will ``feel'' a well defined 
potential along its path. 
If, in addition, $\sigma_{x\nu}$ is small in comparison with the oscillation 
length and the baseline $L$, then one can consider neutrinos as pointlike 
particles. In this case, the potential can vary both in space and time, but at 
any point $\vec{x}$ on the neutrino trajectory only the value of the potential 
at the time $t$ satisfying $\vec{x}=\boldsymbol{v}_\nu t$ will play a role,  
so that $V(t, \vec{x})=V(|\vec{x}|/|\boldsymbol{v}_\nu|, \vec{x})\equiv 
V(\vec{x})$. 

As was discussed in section~{\ref{sec:prob}, the coherent neutrino 
production and detection conditions (\ref{eq:Condit1}) and (\ref{eq:Condit2})
are crucial for the possibility to define the oscillation probability as a 
production- and detection-independent quantity. If these conditions are 
not obeyed, one would have to deal instead with the rate of the overall 
neutrino production-propagation-detection process (\ref{eq:totrate}).
The quantity $\hat{\cal F}$ that enters into this equation is related 
to $\hat{F}$ by eq.~(\ref{eq:mattF}), while $\hat{F}$ should be found as  
the solution of eq.~(\ref{eq:schroed0}) with the boundary condition 
(\ref{eq:boundary}). 
Flavour transitions are then, in general, not directly described by the 
standard neutrino evolution equation in matter (\ref{eq:Sch2}). There are, 
however, exceptions from this rule. First, if the detection coherence 
condition (\ref{eq:Condit1}) is satisfied, the amplitude of the overall 
neutrino production-propagation-detection process satisfies the standard 
evolution equation (\ref{eq:Sch2}), supplemented by the boundary condition 
(\ref{eq:boundary3}). This takes place even if the production coherence 
condition (\ref{eq:Condit1}) is not obeyed and so the oscillation 
amplitude cannot be defined. Second, as shown in Appendix D, in the special 
case of adiabatic neutrino propagation the amplitude of the overall process 
satisfies the standard evolution equation (\ref{eq:Sch2}) even when both 
neutrino production and detection processes are not coherent. The boundary 
condition for the oscillation amplitude is given in this case by 
eq.~(\ref{eq:boundary2}).

Are there any situations in which the coherence conditions for neutrino 
production or detection (\ref{eq:Condit1}), (\ref{eq:Condit2}) are 
violated and therefore the oscillation amplitude satisfying 
the standard evolution equation (\ref{eq:Sch2}) cannot be defined?
As we shall see, this may only be possible for large values of the neutrino 
mass squared differences $\Delta m^2$, which would imply the existence of 
relatively heavy sterile neutrino states. 

Production and detection coherence conditions (\ref{eq:Condit1}) and  
(\ref{eq:Condit2}) actually require that the neutrino production and 
detection regions be small in comparison with the neutrino oscillation 
length (they can therefore be also called the localization conditions).  
Let us consider for simplicity a 2-flavour oscillation problem and discuss 
first neutrino production coherence. The production coherence condition 
(\ref{eq:Condit1}) can then be written as 
\be
\sqrt{\bigg(\frac{\Delta m^2}{2E}\cos 2\theta_0-V(\vec{x}_P)\bigg)^2+
\bigg(\frac{\Delta m^2}{2E}\bigg)^2\sin^2 2\theta_0} \ll \sigma_{pP}\,,
\label{eq:Condit1a}
\ee
where $\theta_0$ is the mixing angle in vacuum. Let us first consider the 
case when the neutrino potential at the production point dominates 
over 
the neutrino kinetic energy difference, i.e. $|V(\vec{x}_P)|\gtrsim 
\Delta m^2/(2E)$. Production coherence condition (\ref{eq:Condit1a}) is then  
violated when 
\be
G_F N_e(\vec{x}_P)\,\gtrsim\, \sigma_{pP}\,\gtrsim\, 1/\sigma_{xP}\,.
\label{eq:Condit1b}
\ee
Assume that the mean distance between the particles of the matter in  
the neutrino production region is $r_0$. Then we have $N_e(\vec{x}_P)\sim 
1/r_0^3$, $\sigma_{xP}\lesssim r_0$, and eq.~(\ref{eq:Condit1b}) requires 
$r_0^2\lesssim G_F$, or $r_0\lesssim 6\cdot 10^{-17}$ cm. This corresponds 
to extremely high densities, exceeding the nuclear density by about ten orders 
of magnitude. Such densities are only attainable in the very early universe,
when neutrino oscillations are irrelevant. 
 
Next, let us consider the opposite situation, $|V(\vec{x}_P)|\ll 
\Delta m^2/(2E)$. Production coherence condition (\ref{eq:Condit1a}) is then  
violated if 
\be
\frac{\Delta m^2}{2E}\,\gtrsim\, \sigma_{pP}\,. 
\label{eq:Condit1c}
\ee
Consider, e.g., neutrinos produced in an accelerator experiment in decays 
of pions of speed $v_\pi$ inside a decay tunnel of length $l_p$. It has 
been shown in \cite{HS,AHS} that in this case the production coherence 
condition is violated 
when 
\be 
\frac{\Delta m^2}{2E}l_p \gtrsim 1\,,
\quad(\Gamma l_p/v_\pi \ll 1)\,;\qquad\qquad
\frac{\Delta m^2}{2E \Gamma}v_\pi \gtrsim 1 
\quad(\Gamma l_p/v_\pi \gg 1)\,,
\label{eq:Condit1d}
\ee
where $\Gamma$ is the pion decay width in the laboratory frame. In the case 
of relatively short decay tunnels ($l_p\ll l_{decay}=v_\pi/\Gamma$) condition 
(\ref{eq:Condit1d}) yields $l_p \gtrsim 2E/\Delta m^2=l_{\rm osc}^{vac}/2\pi$, 
where $l_{\rm osc}^{vac}$ is the vacuum oscillation length. Thus, in this case 
production coherence is violated when the length of the decay tunnel is 
comparable with the neutrino oscillation length. 

The opposite case of relatively long decay tunnels, $l_p\gg  l_{decay}=
v_\pi/\Gamma$, is, however, of more practical interest, since in this case 
most pions decay before being absorbed by the wall at the end of the tunnel.  
In this case we have to use the second inequality in (\ref{eq:Condit1d}), 
which yields $\Delta m^2\gtrsim 1~{\rm eV}^2$. Such values of $\Delta m^2$ 
are currently widely discussed in connection with possible existence of light 
sterile neutrinos \cite{SNAC11,white}. 

Let us now briefly discuss possible detection coherence violation. As follows 
from our discussion above, one can concentrate on the case $|V(\vec{x}_D)|\ll 
\Delta m^2/(2E)$. Detection coherence condition (\ref{eq:Condit2}) is 
then violated provided that 
\be
\frac{\Delta m^2}{2E}\,\gtrsim\, \sigma_{pD}\,\gtrsim\,\frac{1}{\sigma_{xD}}\,, 
\label{eq:Condit2a}
\ee
similarly to (\ref{eq:Condit1c}). Let the average distance between the 
particles in the detector be $r_0$. Then $\sigma_{xD}\lesssim r_0$, and 
condition (\ref{eq:Condit2a}) requires $\Delta m^2/(2E)\gtrsim r_0^{-1}$. For 
matter of normal density $r_0\sim 10^{-9}$ cm, and for neutrinos in the 
MeV range we find that condition (\ref{eq:Condit2a}) requires
$\Delta m^2\gtrsim (100$ keV)$^2$. 

To summarize, we presented a consistent treatment of neutrino oscillations 
in non-inform matter within a QFT framework. We have found that 
the oscillation amplitude can be sensibly defined and can be extracted 
from the amplitude of the overall neutrino production-propagation-detection 
process if neutrinos are ultra-relativistic, matter density varies 
little over the distances of order of the 
sizes of the production and detection regions of individual neutrinos, 
and the neutrino production and detection processes are coherent. By the 
latter we mean that different matter eigenstates composing the flavour states 
are emitted and absorbed coherently. In this case the oscillation amplitude 
satisfies the standard evolution equation (\ref{eq:Sch2}). 
Otherwise one has to consider instead the probability of 
the overall process, given in eq.~(\ref{eq:totrate}). Production coherence 
can be violated e.g.\ in the case of neutrinos produced in decays of free 
pions provided that sterile neutrinos with $\Delta m^2\gtrsim 1$ eV$^2$ exist 
and this mass squared difference plays a role in the flavour transitions of 
interest. For detection processes in matter of normal density (a few g/cm$^3$) 
one can expect coherence violation for $\Delta m^2/(2E)\gtrsim 
(100$ keV)$^2$/MeV.

\vspace*{2mm}
The authors are grateful to Georg Raffelt and Alexei Smirnov for very 
useful discussions. 

\appendix
\renewcommand{\theequation}{\thesection\arabic{equation}}
\appsection
\renewcommand{\thesection}{\Alph{section}}
\section*{Appendix \Alph{section}: 
Neutrino propagator in non-uniform \\ matter}

Here we briefly describe the calculation of the neutrino propagator in 
non-uniform matter in the Dirac and Majorana neutrino cases. In the 
Dirac case our treatment closely follows that of \cite{cardchung}, 
the main difference being that we allow the neutrino mass matrix $M$ 
to be an arbitrary non-singular matrix, whereas in \cite{cardchung} it was 
assumed to be hermitian. 

The coordinate-space neutrino propagator in matter satisfies 
eq.~(\ref{eq:Dir1}). We assume that the matter-induced neutrino 
potential $V^\mu$ is the function of the coordinate $\vec{x}$ along the 
neutrino trajectory but is time independent: $V^\mu=V^\mu(\vec{x})$. 
The neutrino propagator $S_{\beta\alpha}(x, x')$ then 
depends on the times $t$ and $t'$ only through their difference, but on the 
spatial coordinates $\vec{x}$ and $\vec{x}'$ separately: 
$S_{\beta\alpha}(x, x')=S_{\beta\alpha}(t-t';\vec{x},\vec{x}')$. It is 
convenient to introduce the neutrino propagator in the mixed 
energy-coordinate representation $S_{\beta\alpha}(E;\vec{x}, \vec{x}')$, 
which is related to $S_{\beta\alpha}(t-t';\vec{x}, \vec{x}')$ through 
\be
S_{\beta\alpha}(t-t';\vec{x},\vec{x}')=\int \frac{dE}{2\pi}\,
e^{-iE(t-t')}\,S_{\beta\alpha}(E;\vec{x},\vec{x}')\,. 
\label{eq:propagator1}
\ee
The inverse transformation is given by the first equality in 
eq.~(\ref{eq:S2}). 

{}From now on, we will distinguish between the Dirac and Majorana neutrino 
cases.

\subsection{\label{sec:AppDirac}Dirac neutrino propagator}

In this case one has to set $P=P_L$ in eq.~(\ref{eq:Dir1}). 
Omitting the flavour indices to simplify the notation and writing 
$S(E;\vec{x}, \vec{x}')$ in the block-matrix form (\ref{eq:S}), 
from eq.~(\ref{eq:Dir1}) we find 
\be 
\begin{pmatrix} -M^\dag & E+i\boldsymbol{\sigma}\cdot\boldsymbol{\nabla} \\ 
E-i\boldsymbol{\sigma}\cdot\boldsymbol{\nabla}-V^0-\vec{V}\cdot
\boldsymbol{\sigma} 
& -M \end{pmatrix}
\begin{pmatrix} S_{LL} & S_{LR} \\ S_{RL} & S_{RR} 
\end{pmatrix}=\delta^{3}(\vec{x}-\vec{x}')\begin{pmatrix} \mathds{1} & 0 
\\ 0 & 
\mathds{1}\end{pmatrix}\,. 
\label{eq:ChDirac} 
\ee 
We will only need the $S_{LR}$ block matrix of the neutrino propagator. 
{}From~(\ref{eq:ChDirac}) we obtain a system of two coupled 
equations for $S_{LR}$ and \vspace*{1.0mm} $S_{RR}$:
\begin{align}
-&M^\dag 
S_{LR}(E;\vec{x},\vec{x}')+(E+i\boldsymbol{\sigma}\cdot\boldsymbol{\nabla}) 
S_{RR} (E;\vec{x}, \vec{x}')=0\,,
\label{eq:1st}
\\[0.2em]
&[(E-i\boldsymbol{\sigma}\cdot\boldsymbol{\nabla})-V^0(\vec{x})-\vec{V}
(\vec{x})\cdot\boldsymbol{\sigma}]S_{LR}(E;\vec{x}, \vec{x}')-M 
S_{RR}(E;\vec{x}, \vec{x}')=
\delta^{3}(\vec{x}-\vec{x}')\,.
\label{eq:2nd}
\end{align}
\noindent
Next, define 
$J(E;\vec{x}, \vec{x}')\equiv (M^\dag)^{-1} S_{RR}(E;\vec{x}, \vec{x}')$. 
Eq.~(\ref{eq:1st}) then gives 
\be 
S_{LR}(E;\vec{x}, \vec{x}')=(E+i\boldsymbol{\sigma}\cdot 
\boldsymbol{\nabla})J(E; \vec{x}, \vec{x}')\,. 
\label{eq:SundJ1} 
\ee 
Substituting this into eq.~(\ref{eq:2nd}), we obtain the equation for 
$J(E;\vec{x},\vec{x}')$:
\begin{align}
\big\{E^2+\boldsymbol{\nabla}^2-MM^\dag-E V^0(\vec{x})-i\vec{V}(\vec{x})\cdot 
\boldsymbol{\nabla}-i\boldsymbol{\sigma}\cdot[V^0(\vec{x})
\boldsymbol{\nabla}&-iE\vec{V}(\vec{x})+i\vec{V}(\vec{x})\times
\boldsymbol{\nabla}]\big\}\nonumber \\[0.3em] &\times 
J(E;\vec{x},\vec{x}')=\delta^3(\vec{x}-\vec{x}')\,.
\label{eq:main}
\end{align}
We shall seek the solution of this equation in the form
\be
J(E;\vec{x},\vec{x}')=-\frac{e^{i|E||\vec{x}-\vec{x}'|}}
{4\pi|\vec{x}-\vec{x}'|}F(E;\vec{x},\vec{x}')\,.
\label{eq:JundF}
\ee
With this ansatz,  
\be
\boldsymbol{\nabla}J=-\frac{2|E|e^{i|E||\vec{x}-\vec{x}'|}}{4\pi|\vec{x}-
\vec{x}'|} \Bigg[\frac{i\vec{\hat{r}}}{2}F+\frac{1}{2|E|}\boldsymbol{\nabla}F-
\frac{\vec{\hat{r}}}{2|E||\vec{x}-\vec{x}'|}F \Bigg],
\label{eq:auxil2}
\ee
\be
(\boldsymbol{\nabla}^2+E^2)J=
\delta^3(\vec{x}-\vec{x}')F-\frac{2|E|e^{i|E||\vec{x}-\vec{x}'|}}{4\pi|\vec{x}
-\vec{x}'|}\Bigg[\frac{1}{2|E|}\boldsymbol{\nabla}^{2}F
+i(\vec{\hat{r}}\cdot\boldsymbol{\nabla}F)-\frac{1}{|E||\vec{x}-\vec{x}'|}
(\vec{\hat{r}}\cdot\boldsymbol{\nabla} F)\Bigg],
\label{eq:auxil1}
\ee
where $\vec{\hat{r}}\equiv (\vec{x}-\vec{x}')/|\vec{x}-\vec{x}'|$. 
Requiring the first term on the right hand side of (\ref{eq:auxil1}) to 
cancel the $\delta$-function in eq.~(\ref{eq:main}) gives the boundary 
condition for $F$:
\be 
F_{\beta\alpha}(E;\vec{x}, \vec{x}') 
\big|_{\vec{x}\rightarrow \vec{x}'}=\delta_{\beta\alpha}\,, 
\label{eq:boundary} 
\ee 
where we have restored the flavour indices $\alpha$ and $\beta$.
Since in neutrino oscillation experiments one deals with macroscopic 
distances, we are primarily interested in well separated $\vec{x}$ and 
$\vec{x}'$. 
This means $|E||\vec{x}-\vec{x}'|\gg 1$, so that the last terms in 
eqs.~(\ref{eq:auxil1}) and (\ref{eq:auxil2}) can be neglected. 
Eq.~(\ref{eq:main}) then becomes 
\begin{align}
i(\vec{\hat{r}}\cdot\boldsymbol{\nabla}F)&+\frac{1}{2|E|}\boldsymbol{\nabla}^2
F-\frac{1}{2|E|}[MM^\dag+EV^0-|E|(\vec{\hat{r}}\cdot\vec{V})
\nonumber\\
&-\boldsymbol{\sigma}\cdot
(V^0|E|\vec{\hat{r}}-E\vec{V}+i|E|\vec{V}\times\vec{\hat{r}})]F+{\cal O}
\left(\frac{|V^\mu||\boldsymbol{\nabla}F|}{|E|}\right)=0\,,
\label{eq:almostS}
\end{align}
Since in all situations of practical interest matter-induced neutrino 
potentials are by far much smaller than neutrino energy,
\be
|V^\mu| \ll |E|\,,
\label{eq:small}
\ee
the last term in (\ref{eq:almostS}) can be neglected in comparison 
with the first term.  
Choosing the $z$-axis of the coordinate system along $\vec{\hat{r}}$, 
one can then rewrite eq.~(\ref{eq:almostS}) as 
\be
i(\vec{\hat{r}}\cdot\boldsymbol{\nabla}F)+\frac{1}{2|E|}\boldsymbol{\nabla}^2F-
\frac{1}{2|E|}D(E, \vec{x})F=0\,,
\label{eq:schr2}
\ee
where 
\be
D(E,\vec{x})=\left(
\begin{array}{ll} 
MM^\dag+(E-|E|)(V^0+V^3) &~~ (E+|E|)(V^1-iV^2)\\[0.6em] 
(E-|E|)(V^1+iV^2) &~~ MM^\dag+(E+|E|)(V^0-V^3) 
\vspace*{0.5mm}
\end{array}\right).
\label{eq:Dmatrix}
\ee
Let us now distinguish three cases: 
(1) $|\boldsymbol{\nabla}F|\gg \epsilon F$, 
where $\epsilon$ is the largest eigenvalue of the matrix 
$D(E,\vec{x})/2|E|$;
(2) $|\boldsymbol{\nabla}F|\ll \epsilon F$; and 
(3) $|\boldsymbol{\nabla}F| \sim \epsilon F$.
The first case (in which the third term in eq.~(\ref{eq:schr2}) can 
be neglected) is of no interest to us because it corresponds to 
the kinematic region in which neutrinos essentially do not oscillate. 
In the second case the three terms in eq.~(\ref{eq:schr2}) cannot 
balance each other, i.e.\ this equation cannot be satisfied. This 
immediately follows from (\ref{eq:small}) and the condition 
$\Delta m^2 \ll E^2$, where $\Delta m^2=\max\{\Delta m_{ik}^2\}$.%
\footnote{\footnotesize 
Obviously, only mass squared differences and 
not the absolute neutrino  masses play a role in neutrino oscillations. Technically, this can be proven 
by subtracting from $D(E,\vec{x})$ the matrix 
$m_i^2 \left(
\begin{array}{cc} {\mathbbm 1} & 0 \\
0 & {\mathbbm 1}\end{array}
\right)$, 
where $m_i$ is any neutrino mass eigenvalue 
and ${\mathbbm 1}$ is the unit matrix in the flavour space, and 
rephasing 
$F$ accordingly. }
Thus, the only case of interest to us is the third one. It is easy to see that 
in this case the second term in eq.~(\ref{eq:schr2}) is negligibly small 
compared to the other two and so can be omitted provided that neutrinos 
are relativistic, the components of the neutrino potential $V^\mu(\vec{x})$ 
satisfy eq.~(\ref{eq:small}), and in addition 
\be
\Big|\frac{\boldsymbol{\nabla}V^\mu}{V^\mu}\Big|\ll |E|\,.
\label{eq:condit1}
\ee
Note that this condition 
requires that 
the potential change little over the distances of order of the neutrino 
de Broglie wavelength. 
Under the above conditions eq.~(\ref{eq:schr2}) reduces to  
\be
i(\vec{\hat{r}}\cdot\boldsymbol{\nabla}F)-\frac{1}{2|E|}D(E, \vec{x})F=0\,. 
\label{eq:schr3}
\ee

Let us now concentrate on the case of neutrinos, $E>0$ (the antineutrino 
case can be studied similarly). From $|\boldsymbol{\nabla}F| \sim \epsilon F$ 
we find that the second term on the right hand side of eq.~(\ref{eq:auxil2}) 
is much smaller than the first one; we have already established that the third 
term in this equation is negligible. Thus, $(E+i\boldsymbol{\sigma}\cdot 
\boldsymbol{\nabla})J\approx E(1-\boldsymbol{\sigma}\cdot 
\vec{\hat{r}})J=E(1-\sigma_3)J$. From eq.~(\ref{eq:SundJ1}) we then find 
\be
(S_{LR})_{22}=2E J_{22}\,,
\label{eq:SundJ2}
\ee 
with all the other spinor components of   
$S_{LR}$ being zero. This result plays an important role in our  
calculations, since we need to deal only with one component of the 
neutrino propagator, and this simplifies our consideration significantly. 
Eq. (\ref{eq:JundF}) relating $S_{LR}$ and $F$ then implies that the only 
relevant spinor component of $F$ is $F_{22}$. 

Next, we note that for $E>0$ eq.~(\ref{eq:Dmatrix}) can be rewritten as   
\be
D(E,\vec{x})=\left(
\begin{array}{ll} 
MM^\dag &~~ 2E(V^1-iV^2)\\[0.6em] 
0 &~~ MM^\dag+2E(V^0-V^3) 
\vspace*{0.5mm}
\end{array}\right).
\label{eq:Dmatrix2}
\ee
The fact that $D_{21}=0$ means, in particular, that the equation for the 
spinor component $F_{22}$ in (\ref{eq:schr3}) decouples, i.e. does not 
contain any other components of $F$. 

Denoting the 22-components of $S_{LR}$, $J$ and $F$ as $(S_{LR})_{22}\equiv 
\hat{S}$, $J_{22}\equiv \hat{J}$ and $F_{22}=\hat{F}$, we finally obtain from 
(\ref{eq:SundJ2}), (\ref{eq:JundF}), (\ref{eq:schr3}) and (\ref{eq:Dmatrix2})
\be 
\hat{S}_{LR}(E;\vec{x},\vec{x}')=-2E\frac{e^{i|E||\vec{x}-\vec{x}'|}}
{4\pi|\vec{x}-\vec{x}'|}\hat{F}(E;\vec{x},\vec{x}')\,,
\label{eq:propF1a}
\ee
where $\hat{F}$ satisfies the Shr\"odinger-like equation  
\be
i\frac{d}{dx}\hat{F}=\left[\frac{M\!M^\dag}{2|E|}+V(\vec{x})\right]\hat{F}\,.
\label{eq:schroed1}
\ee
Here $\frac{d}{dx}$ as the directional derivative along $\vec{\hat{r}}$: 
$\frac{d}{dx}\equiv\vec{\hat{r}}\cdot \boldsymbol{\nabla}$.
The potential $V(\vec{x})$ is defined as  
\be
V(\vec{x})\equiv V^0(\vec{x})-V^3(\vec{x})\simeq V^0(\vec{x})-
\boldsymbol{v}_\nu\!\cdot\! 
\vec{V}(\vec{x})\,,
\label{eq:pot}
\ee
where $\boldsymbol{v}_\nu$ ($|\boldsymbol{v}_\nu|\simeq 1$) is the neutrino 
velocity vector, and the last equality is valid for an arbitrary orientation 
of the coordinate axes. Eq.~(\ref{eq:schroed1}) actually describes both the 
neutrino and antineutrino cases; the potential for antineutrinos is obtained 
from that for neutrinos by flipping the sign of the latter (except in 
CP-symmetric or nearly CP-symmetric media, see Appendix B).%
\footnote{The antineutrino case is studied quite analogously to the 
neutrino one. In that case one has to replace $E\to -E$, $\vec{\hat{r}}\to 
-\vec{\hat{r}}$. 
The only non-vanishing spinor components of $S_{LR}$ and $J$ are 
$S(_{LR})_{11}$ and $J_{11}$, and the only relevant component of 
$F$ is $F_{11}$, which satisfies the same eq.~(\ref{eq:schroed1}) with 
the potential being negative of the neutrino potential (except in media 
with equal or almost equal numbers of particles and antiparticles).      
The right-handed antineutrino spinors in the momentum space 
are $v_R(p)=(\sqrt{2p_0},\;0)^T$, i.e.\ only their upper components are  
non-zero.}

Note that the factor $\frac{e^{i|E||\vec{x}_1-\vec{x}_2|}}{4\pi|\vec{x}_1
-\vec{x}_2|}$ in expression (\ref{eq:propF1a}) for $\hat{S}_{LR}$ is a 
fast varying function of the coordinates, which changes significantly over 
distances of order of the neutrino de Broglie wavelength $E^{-1}$, whereas 
eq.~(\ref{eq:schroed1}) actually means that the factor $\hat{F}$ is a slowly 
varying function of $x$, which changes significantly over the distances of 
order $\min\{E/\Delta m^2,\, |V^0|^{-1},\, |\vec{V}|^{-1}\}$, i.e. of 
order of neutrino oscillation length in matter.  
 
For a known matter-induced potential, eqs.~(\ref{eq:propF1a}) and 
(\ref{eq:schroed1}) together with the boundary condition (\ref{eq:boundary}) 
fully determine the neutrino propagator 
$\hat{S}_{LR}(E,\vec{x},\vec{x}')$. 

\subsection{\label{sec:AppMajor}Majorana neutrino propagator}

Recall that for Majorana neutrinos we use the Feynman rules in which 
propagators and vertices do not contain explicitly the charge-conjugation 
matrix \cite{gates,denner}. 
Let is first discuss the choice $P=-\gamma_5$ in eqs.~(\ref{eq:DirProp}) and 
(\ref{eq:Dir1}).  
For Majorana neutrinos  
the 4-component field $\nu=\nu_L+\nu_R$ can be written 
as $\nu=\nu_L+(\nu_L)^c$, where the superscript $c$ means charge conjugation. 
In other words, in this case right-handed neutrinos are antiparticles 
of left-handed ones, and so they participate in the standard weak 
interactions. The matter-induced potential $V^\mu$ enters the equations of 
motion of the right-handed and left-handed fields with opposite signs. 
The choice $P=-\gamma_5$ in eqs.~(\ref{eq:DirProp}) and (\ref{eq:Dir1}) 
in the Majorana neutrino case then follows from the relations $-\gamma_5 
\nu_L=\nu_L$ and $-\gamma_5 \nu_R=-\nu_R$. 

Consider now eq.~(\ref{eq:Dir1}). 
Using, as before, the block-matrix form for the neutrino propagator in 
the mixed coordinate-energy representation $S(E;\vec{x},\vec{x}')$ , we 
arrive at the equation 
\be 
\begin{pmatrix} -M^* & E+i\boldsymbol{\sigma}\cdot\boldsymbol{\nabla} 
+V^0-\vec{V}\cdot\boldsymbol{\sigma} \\ 
E-i\boldsymbol{\sigma}\cdot\boldsymbol{\nabla}-V^0-\vec{V}\cdot
\boldsymbol{\sigma} & -M \end{pmatrix}
\begin{pmatrix} S_{LL} & S_{LR} \\ S_{RL} & S_{RR} 
\end{pmatrix}=\delta^{3}(\vec{x}-\vec{x}')\begin{pmatrix} \mathds{1} & 0 
\\ 0 & 
\mathds{1}\end{pmatrix}\,. 
\label{eq:ChDirac2} 
\ee 
It differs from eq.~(\ref{eq:ChDirac}) by the presence of the 
potential-dependent term in the 12-entry of the first matrix on the left hand 
side. Since in the case of Majorana neutrinos $M$ is in general complex 
symmetric, we replaced $M^\dag$ by $M^*$.
{}From~(\ref{eq:ChDirac2}) we obtain a system of two coupled 
equations for $S_{LR}$ and $S_{RR}$:
\begin{align}
-&M^*
S_{LR}(E;\vec{x},\vec{x}')+(E+i\boldsymbol{\sigma}\cdot\boldsymbol{\nabla} 
+V^0(\vec{x})-\vec{V}(\vec{x})\cdot\boldsymbol{\sigma})
S_{RR} (E;\vec{x}, \vec{x}')=0\,,
\label{eq:1stM}
\\[0.2em]
&[(E-i\boldsymbol{\sigma}\cdot\boldsymbol{\nabla})
-V^0(\vec{x})-\vec{V}(\vec{x})\cdot\boldsymbol{\sigma}
]S_{LR}(E;\vec{x}, \vec{x}')-M 
S_{RR}(E;\vec{x}, \vec{x}')=
\delta^{3}(\vec{x}-\vec{x}')\,.
\label{eq:2ndM}
\end{align}
\noindent
Note that eq.~(\ref{eq:2ndM}) coincides with (\ref{eq:2nd}), whereas 
eq.~(\ref{eq:1stM}) differs from (\ref{eq:1st}) by an extra 
potential-dependent term in the coefficient of $S_{RR}(E;\vec{x},\vec{x}')$. 
Next, we define, as before,  
$J(E;\vec{x}, \vec{x}')\equiv (M^*)^{-1} S_{RR}(E;\vec{x}, \vec{x}')$. 
Eq.~(\ref{eq:1stM}) then gives 
\be 
S_{LR}(E;\vec{x}, \vec{x}')=\big[E+i\boldsymbol{\sigma}\cdot 
\boldsymbol{\nabla}+V^0(\vec{x})-\vec{V}(\vec{x})\cdot\boldsymbol{\sigma}\big]
J(E; \vec{x}, \vec{x}')\,. 
\label{eq:SundJ2A} 
\ee 
For relativistic neutrinos $(E+i\boldsymbol{\sigma}\cdot \boldsymbol{\nabla})
J(E; \vec{x},\vec{x}')\approx 2E J(E; \vec{x}, \vec{x}')$,  
therefore, under condition (\ref{eq:small}) one can neglect the 
term $V^0(\vec{x})-\vec{V}(\vec{x})\cdot\boldsymbol{\sigma}$ 
in~(\ref{eq:SundJ2A}).%
\footnote{Note that we cannot neglect the similar term in eq.~(\ref{eq:2ndM}) 
because the coefficient of $S_{LR}(E;\vec{x},\vec{x}')$ in this equation 
contains $(E-i\boldsymbol{\sigma}\cdot\boldsymbol{\nabla})$ rather than 
$(E+i\boldsymbol{\sigma}\cdot\boldsymbol{\nabla})$.}  
Eq.~(\ref{eq:SundJ2A}) then reduces to eq.~(\ref{eq:SundJ1}). Since 
eqs.~(\ref{eq:2ndM}) and (\ref{eq:2nd}) coincide, we find that the 
neutrino propagator is still given by (\ref{eq:propF1a}), 
where $\hat{F}(E,\vec{x},\vec{x}')$ satisfies eq.~(\ref{eq:schroed1}) 
with the boundary condition~(\ref{eq:boundary}). 

Thus, for propagation of relativistic neutrinos in matter with potential  
satisfying $|V^\mu(\vec{x})|\ll |E|$ the propagator of Majorana neutrinos 
coincides with that of Dirac neutrinos.

\appsection
\renewcommand{\thesection}{\Alph{section}}
\section*{Appendix \Alph{section}: 
Matter-induced neutrino potentials}

We summarize here the expressions for the potentials of relativistic 
neutrinos caused by coherent forward scattering of neutrinos on background 
particles. For definiteness, we concentrate on the Dirac neutrino case; 
the potentials for Majorana neutrinos are the same, the only difference 
being that what we call antineutrinos in the Dirac case are just 
right-handed neutrino components in the Majorana case. 

Neutrino interact with matter through the charged current (CC) and neutral 
current (NC) interactions mediated by $W^\pm$ and $Z^0$ bosons, 
respectively. As we shall show, the effective Lagrangian of neutrino 
interaction with matter can be written as 
\be
{\cal L}_{int}=-\bar{\nu}\,(\gamma_\mu V^\mu) P_L\, \nu\,,
\label{eq:L}
\ee
where 
the matrix of matter-induced neutrino potentials $V^\mu$ is diagonal 
in the flavour basis%
\footnote{Except in media containing neutrino backgrounds, see below.}
and is the sum of the CC and NC contributions: 
$V^\mu=V_{\rm CC}^\mu+V_{\rm NC}^\mu$. Adding ${\cal L}_{int}$ to the free 
neutrino Lagrangian and making use of the standard Euler-Lagrange formalism 
to derive the neutrino equation of motion, one arrives at 
eq.~(\ref{eq:Dir1}) for the neutrino propagator in matter.   

We shall now concentrate on the potentials $V_{\rm CC}^\mu$ and 
$V_{\rm NC}^\mu$. We will be assuming (except in eq.~(\ref{eq:contrib1})
below) that the energies of neutrinos and particles of the medium are small 
compared to the $W$-boson mass $m_W$. 
In an ordinary matter with no muons or tauons present, only electron 
neutrinos experience CC interactions, which are due to their scattering on the 
electrons of the medium. The effective Lagrangian of this   
interaction is  
\be
{\cal L}_{\rm CC}=-\frac{G_F}{\sqrt{2}}[\bar{\nu}_{e}(x)\gamma^\mu(1-\gamma_5) 
e(x)][\bar{e}(x) \gamma_\mu(1-\gamma_5) \nu_e(x)]\,,
\label{eq:LCC1}
\ee
where $G_F$ is the Fermi constant. We then employ the Fierz transformation 
to permute the neutrino field with the electron one and take the expectation 
value of the electron current over the state of the medium. This gives 
\be
[{\cal L}_{\rm CC}]_{\nu_e}=-\bar{\nu}_{e}\gamma_\mu \big[(V_e)_{\rm CC}^\mu
(x)\big]P_L\nu_e\,,
\label{eq:LCC2}
\ee
where \cite{Wolf,lls}  
\be
(V_e)_{\rm CC}^\mu (x)=\sqrt{2}\, G_F\langle\bar{e}(x)\gamma^\mu
(1-\gamma_5) e(x)\rangle\,.
\label{eq:V1}
\ee
Here $\langle ... \rangle$ means the average over the state of the medium, 
and we have taken into account that for relativistic left-handed neutrinos 
$(1-\gamma_5)\nu_L\approx 2\nu_L$. By making use of the solutions of the 
Dirac equation for electrons, for the expectation values of 
the components of the electron current we find 
\begin{align}
& \langle\bar{e}(x)\gamma^0 e(x)\rangle = N_{e}(x)\,,\quad 
\langle\bar{e}(x)\gamma^i e(x)\rangle = N_e(x)v_{e}^i(x)\,, \quad
\langle\bar{e}(x)\gamma^0\gamma_5 e(x)\rangle =N_e(x) 
\langle \boldsymbol{\sigma}_e\!\cdot\! \vec{v}_{e}\rangle\,, 
\nonumber \\[0.3em]
& \langle\bar{e}(x)\gamma^i\gamma_5 e(x)\rangle = N_e(x)\big[
m_e \langle \sigma^i/E_e \rangle + 
\langle [E_e/(E_e+m_e)]
v_{e}^i(\boldsymbol{\sigma}_e\!\cdot\!
\vec{v}_{e})\rangle
\big] \,, 
\label{eq:V2}
\end{align}
where $N_e(x)$ is the electron number density, $v_e^i(x)$ is the $i$th 
component of the electron velocity, and 
$\sigma_e^i$ are the electron Pauli matrices ($i=1,2,3$). Note that the 
expectation  values of all the components of the axial-vector current 
vanish in a medium with unpolarized electrons. For such a medium from 
(\ref{eq:V1}) and (\ref{eq:V2}) we obtain 
\be
(V_e)_{\rm CC}^0 (x)=\sqrt{2}\, G_F N_{e}(x)\,,\qquad
(V_e)_{\rm CC}^i (x)=\sqrt{2}\, G_F N_e(x) v_{e}^i(x)\,, 
\label{eq:V3}
\ee
The CC contribution to the expression $V=V^0-\vec{v}_\nu\!\cdot\!\vec{V}$ that 
enters into eqs.~(\ref{eq:schroed0}) and (\ref{eq:schroed1}) \vspace*{1mm}
is then 
\be
(V_e)_{\rm CC}=\sqrt{2}\, G_F N_{e}
(1-v_{e}\cos\theta_{e\nu})\,, 
\vspace*{1mm}
\label{eq:V4}
\ee
where $\theta_{e\nu}$ is the angle between the momenta of the electron and the 
neutrino. For media with electrons at rest or non-relativistic electrons 
\mbox{($v_e\ll 1$)} the spatial components of the CC potential can be 
neglected, and one obtains $V_{\rm CC}\simeq V_{\rm CC}^0=
\sqrt{2}G_F N_e(x)$. This is the expression for the neutrino potential which 
is relevant e.g.\ for neutrino oscillations in the sun and inside the 
earth. It should be noted, however, that during supernova collapse   
or in rotating neutron stars bulk matter velocities may be substantial,   
leading to non-negligible net fluxes. In those cases the terms 
in the neutrino potentials that depend on the velocities of background 
particles should be retained.

Consider now NC contributions to the matter-induced neutrino potentials. 
The effective Lagrangian of the NC interaction of $\nu_\alpha$ 
($\alpha=e,\,\mu,\,\tau$) with a fermion $f$ where $f=e,\,p,\,n$ (or a 
background neutrino which may be abundant in supernovae or in the early 
universe) is 
\be
{\cal L}_{\rm NC}=-\frac{G_F}{\sqrt{2}}\,
[\bar{\nu}_{\alpha}(x)\gamma^\mu(1-\gamma_5)\nu_\alpha(x)]\,
[\bar{\psi}_f(x) \gamma_\mu(T_{3Lf}-2 Q_f\sin^2\theta_W)\psi_f(x)]\,.
\label{eq:LNC1}
\ee
Here $Q_f$ and $T_{3Lf}$ are the electric charge of the fermion $f$ and 
the third isospin projection of its left-handed component, respectively, 
and $\theta_W$ is the Weinberg angle.  
Similarly to eq.~(\ref{eq:LCC2}), upon averaging the variables of the fermion 
$f$ over the state of the matter we find 
\be
[{\cal L}_{\rm NC}]_{\nu_\alpha,f}=-\bar{\nu}_{\alpha}\gamma_\mu 
\big[(V_\alpha)_{\rm NC}^\mu (x)\big]_f P_L\,\nu_\alpha\,,
\label{eq:LNC2}
\ee
where 
\be
[(V_\alpha)_{\rm NC}^\mu (x)]_f=\sqrt{2}\, G_F (T_{3Lf}-2 Q_f\sin^2\theta_W)
\,\langle\bar{\psi}_f(x) \gamma^\mu \psi_f(x)\rangle\,.
\label{eq:V5}
\ee
It is important to note that the NC-induced potentials (\ref{eq:V5}) 
do not depend on the neutrino flavour index $\alpha$. That is, they are the 
same for all three active neutrino species ($\nu_e$, $\nu_\mu$ and $\nu_\tau$) 
and vanish for sterile neutrinos. The equality $[(V_e)_{\rm NC}^\mu]_f=
[(V_\mu)_{\rm NC}^\mu]_f=[(V_\tau)_{\rm NC}^\mu]_f$ actually holds only at 
tree level; at one-loop level tiny differences between these potentials arise, 
which are usually irrelevant. They can, however, play some role at extremely 
high densities, e.g.\ in supernovae. We will consider loop-induced NC 
contributions to the potentials below. 

The expectation value $\langle\bar{\psi}_f \gamma^\mu \psi_f\rangle$ can be 
obtained from eq.~(\ref{eq:V2}) by replacing the subscript $e$ by $f$. We will 
assume now that the particles $f$ are unpolarized and have zero mean 
velocities or are non-relativistic and therefore one can keep only the time 
components of the NC-induced neutrino potentials. For the NC contributions of 
neutrino scattering on the electrons, protons and neutrons of the matter we 
then find \cite{notzraff}
\begin{align}
&[(V_\alpha)_{\rm NC} (x)]_e=\sqrt{2}\, G_F N_e(x)\Big(-\frac{1}{2}+
2\sin^2\theta_W\Big),\nonumber \\
& [(V_\alpha)_{\rm NC} (x)]_p=\sqrt{2}\, G_F N_p(x)\Big(\frac{1}{2}-
2\sin^2\theta_W\Big),\nonumber \\
&[(V_\alpha)_{\rm NC} (x)]_n=\sqrt{2}\, G_F \Big(-\frac{N_n(x)}{2}\Big).
\label{eq:V6}
\end{align}
In an electrically neutral matter one has $N_e(x)=-N_p(x)$, so that the 
electron and proton contributions cancel each other, and the only non-zero 
net effect is due to the neutrons. Combining the 
CC and NC contributions to the neutrino potential, in the flavour basis 
$(\nu_e,\,\nu_\mu,\,\nu_\tau,\,\nu_s)$ where $\nu_s$ is a hypothetical 
sterile neutrino, we get for the matrix $V=V_{\rm CC}+V_{\rm NC}$
\be
V=\sqrt{2} G_F\, {\rm diag}\Big(N_e-\frac{N_n}{2},\;-\frac{N_n}{2},\;
-\frac{N_n}{2},\;0 \Big).
\label{eq:V7}
\ee
For antineutrinos the right-hand side of this equality should be multiplied 
by $-1$.

Since one can always add to the effective Hamiltonian in 
eq.~(\ref{eq:Sch2}) any matrix proportional to the unit matrix without 
affecting the oscillation probabilities, 
one can modify the matrix $V$ in (\ref{eq:V7}) according to $V\to V+G_F 
(N_n/\sqrt{2})\!\cdot\!\mathbbm{1}$. This yields 
\be
V=\sqrt{2} G_F\, {\rm diag} \Big(N_e,\;0,\;0,\;\frac{N_n}{2}\Big).
\label{eq:V8}
\ee
For neutrino propagation in normal media, this form is the most often 
used one. It is especially convenient when 
oscillations between only active neutrino species are considered, since 
in this case the matrix $V$ in (\ref{eq:V8}) has only one non-zero element 
$V_e(x)=\sqrt{2}G_F N_e(x)$. Thus, in this case only the CC contribution to 
the neutrino potential affects neutrino oscillations. Note a useful relation 
$\sqrt{2}G_F N_e\simeq 7.63\times 10^{-14} \rho Y_e$ eV, 
where $\rho$ is the matter density in g/cm$^3$ and $Y_e$ is the 
electron fraction (number of electrons per baryon) in matter.  

Loop corrections to the matter-induced neutrino potentials were 
calculated in \cite{botlimmarc}. They differ for neutrinos of different 
flavour due to the differences of the masses of the corresponding 
charged leptons. The most important difference is the one between the 
potentials of $\nu_\tau$ and $\nu_\mu$, since this difference vanishes 
at tree level (see (\ref{eq:V7}) or (\ref{eq:V8})). For a neutral 
unpolarized medium it is 
\be
\Delta V_{\tau\mu}\equiv V_\tau-V_\mu\approx \pm \frac{3}{2\pi^2} G_F^2 
m_\tau^2\Big[(N_p+N_n)\ln\frac{m_W^2}{m_\tau^2}-(N_p+\frac{2}{3}N_n)\Big].
\label{eq:V9}
\ee
Here and below the upper sign always refers to neutrinos and the lower 
one to antineutrinos. 
Note that $\Delta V_{\tau\mu}$ is very small,  
$\Delta V_{\tau\mu}/V_e\sim 5\times 10^{-5}$. However, it may play 
some role at very high densities, in particular, for supernova neutrinos 
\cite{ALS}. One-loop contributions to $\Delta V_{\tau\mu}$ in the 
neutrino backgrounds were calculated in \cite{mprs}.

The above formulas for the matter-induced neutrino potentials apply to 
the ordinary unpolarized matter at zero temperature and with no antiparticles. 
We will now relax these constraints.

\bigskip
\noindent
{\bf Neutrino potentials in hot and dense matter and in neutrino backgrounds}
\vspace*{1.0mm}

\noindent
This case is relevant for the early universe and supernova physics. It was 
studied in refs.~\cite{notzraff,raffsigl}, the results of which we summarize 
here. 
In an electrically 
neutral unpolarized medium consisting in general of electrons, muons, 
$\tau$-leptons, protons, neutrons and their antiparticles with zero 
mean velocities, 
the potential of electron (anti)neutrinos is
\be
V_e=\pm \sqrt{2} G_F \Big[(N_e-N_{\bar{e}})-\frac{1}{2}(N_n-N_{\bar{n}})
\mp\frac{2E}{m_W^2}\big( 
\langle E_e(1+v_e^2/3)\rangle N_e+\langle E_{\bar{e}}(1+v_{\bar{e}}^2/3)
\rangle N_{\bar{e}}\big)
\Big].
\label{eq:V10}
\ee
Here $E$ is the energy of the neutrino, $E_e$ and 
$E_{\bar{e}}$ are those of the electrons and positrons of the medium, 
$v_e$ and $v_{\bar{e}}$ are the electron and positron velocities, 
and $N_{\bar{f}}$ stands for the number density of the antiparticles of $f$. 
All the averages are now taken over the proper thermal distributions of the 
background particles. Note that the first and second terms in the square 
brackets in (\ref{eq:V11}) are the generalizations of the CC and NC 
contributions to $V_e$ discussed above to the case when antiparticles 
are present in matter. The NC-induced term comes only from the neutrino 
scattering on neutrons, since the NC contributions of all charged 
particles cancel in an electrically neutral medium.  
The third term in (\ref{eq:V11}) is due to CC and is rather special. It comes 
from the second-order term in the expansion of the $W^\pm$ propagator in powers 
of $1/m_W^2$. Due to an extra power of $m_W^2$ in the denominator, it is 
negligibly small in an ordinary matter. 
However, it does not vanish in the limit $N_e=N_{\bar{e}}$ and so becomes 
important in a medium with equal (or almost equal) abundances of particles 
and antiparticles, when the contributions of the first two terms are 
negligible. In addition, this term has the same (negative) sign for 
electron neutrinos and antineutrinos.

The last property, as well as the fact that the third term in 
(\ref{eq:V10}) is non-zero for $N_e=N_{\bar{e}}$, can be understood as 
follows. The contribution of the $W$-boson exchange to $\nu_e e$ 
scattering amplitude is proportional to $g^2/[m_W^2-(q-p)^2]\approx 
(g^2/m_W^2)(1-2 q\!\cdot\! p/m_W^2)$, where $q$ and $p$ are 4-momenta of 
the neutrino and of a background electron, and $g$ is the $SU(2)_L$ 
gauge coupling constant. For $\nu_e$ scattering on positrons one has to 
flip the overall sign of this expression and to replace $p\to -p$, so 
that the corresponding contribution to $V_e$ is proportional to 
$-g^2/[m_W^2-(q+p)^2]\approx -(g^2/m_W^2)(1+2 q\!\cdot\! p/m_W^2)$. 
Obviously, the terms $\sim 1/m_W^4$ enter with the same (negative) sign.  
The situation is similar if one goes from neutrinos to antineutrinos, in 
which case one has to replace $g^2 \to -g^2$, $q\to -q$. 
The factor $\langle E_e(1+v_e^2/3)\rangle$ comes from the averaging of 
$E_e(1-v_e\cos\theta_{\vec{q}\vec{p}})^2$ over the angle 
$\theta_{\vec{q}\vec{p}}$ between the momenta of the neutrino and 
the background electron.%
\footnote{One power of $(1-v_e\cos\theta_{\vec{q}\vec{p}})$ is due to the fact 
that for relativistic neutrinos $V=V^0-V^3\simeq V^0(1-v_e\cos\theta_{\vec{q}
\vec{p}})$, while the other power and the factor $E_e$ come from 
$p\!\cdot\!q\simeq E E_e (1-v_e \cos\theta_{\vec{q}\vec{p}})$.}

Another interesting propagator effect takes place at extremely high neutrino 
and/or electron energies. In a CP-symmetric matter with equal electron and 
positron abundances the CC contribution to the matter-induced 
self-energy of $\nu_e$ is proportional to 
\be
g^2\bigg[
\frac{1}{m_W^2+2p\!\cdot\!q}-\frac{1}{m_W^2-2p\!\cdot\!q}\bigg]=
-g^2\,\frac{4p\!\cdot\!q}{m_W^4-4(p\!\cdot\!q)^2}\,,
\vspace*{1mm}
\label{eq:contrib1}
\ee 
where no expansion in powers of $1/m_W^2$ has been done. In the limit 
$(p\!\cdot\!q)^2\ll m_W^4$ the previous results are recovered, whereas we 
see that for $4(p\!\cdot\!q)^2> m_W^4$ the potential changes its sign. 

The potentials of $\nu_\mu(\bar{\nu}_\mu)$ and $\nu_\tau(\bar{\nu}_\tau)$ in 
matter are given by expressions similar to (\ref{eq:V10}), with the index $e$ 
replaced by $\mu$ or $\tau$, respectively. If the medium contains no 
$\mu^\pm$ and $\tau^\pm$, only the neutron contributions to $V_{\mu}$ and 
$V_{\tau}$ (which coincide with the second term in (\ref{eq:V10})) survive.

In a number of applications (e.g. for supernova neutrinos) it is necessary 
to consider neutrino potentials in neutrino backgrounds. Those are due 
to the NC interactions, and they depend on whether the background of the same 
flavour or different flavour neutrinos is considered. In the case of the 
same flavour neutrino background the corresponding contribution to the 
potential of the test neutrino of momentum $\vec{q}$ is  
\begin{align}
\Delta V_\alpha=\sqrt{2} G_F \int\!&\frac{d^3p}{(2\pi)^3}\Big\{\pm
2\big(n_{\nu_\alpha}^L(\vec{p})-n_{\bar{\nu}_\alpha}^L(\vec{p})\big)
(1-\cos\theta_{\vec{q}\vec{p}}) 
\nonumber \\[0.2em]
&-\frac{2E_{\nu_\alpha}(\vec{q})}{m_Z^2}\Big[E_{\nu_\alpha}(\vec{p})
n_{\nu_\alpha}^L(\vec{p})+E_{\bar{\nu}_\alpha}(\vec{p})
n_{\bar{\nu}_\alpha}^L(\vec{p})\Big]
(1-\cos\theta_{\vec{q}\vec{p}})^2\Big\}.
\label{eq:V12}
\end{align}
Here $n_{\nu_\alpha}^L(\vec{p})$ and $n_{\bar{\nu}_\alpha}^L(\vec{p})$ are the 
occupation numbers of the left-handed background neutrinos of flavour 
$\alpha$ and of their antiparticles. The quantity $n_{\nu_\alpha}^L(\vec{p})$ 
is related to the neutrino number density $N_{\nu_\alpha}^L$ through 
\be 
N_{\nu_\alpha}^L=\int\frac{d^3p}{(2\pi)^3}\,n_{\nu_\alpha}^L(\vec{p})\,,
\label{eq:occup}
\ee
and similarly for antineutrinos. 
The origin of the last term in the curly brackets is similar to that of 
the last term in (\ref{eq:V10}), except that it comes from the 
expansion of the $Z$ boson rather than $W$ boson propagator. 
Note that the neutrino 
potential due to the coherent forward scattering on a background 
neutrino vanishes when the velocities of the test and background neutrinos 
are parallel to each other, i.e.\ when $\cos\theta_{\vec{q}\vec{p}}=1$. This 
happens because there is no forward neutrino-neutrino scattering for 
completely relativistic neutrinos moving in the same direction.
If the momentum distribution of the background neutrinos is isotropic, 
then $\langle \cos\theta_{\vec{q}\vec{p}}\rangle=0$ 
and (\ref{eq:V12}) reduces to 
\be
\Delta V_\alpha=\pm 2\sqrt{2} G_F 
\big(N_{\nu_\alpha}^L-N_{\bar{\nu}_\alpha}^L)
-\frac{8E_{\nu_\alpha}(\vec{q})\sqrt{2}G_F}{3m_Z^2}\Big[
\langle E_{\nu_\alpha}\rangle
N_{\nu_\alpha}^L+\langle E_{\bar{\nu}_\alpha}\rangle
N_{\bar{\nu}_\alpha}^L(p)\Big].\\[0.2em]
\label{eq:V12a}
\ee

For a test neutrino in a neutrino background of different flavour one has 
\be
\Delta V_\alpha=\pm\sqrt{2} G_F \int\!\frac{d^3p}{(2\pi)^3}
\big(n_{\nu_\beta}^L(\vec{p})-n_{\bar{\nu}_\beta}^L(\vec{p})\big)
(1-\cos\theta_{\vec{q}\vec{p}}) ~~~\quad\qquad (\beta\ne \alpha)\,.
\label{eq:V13}
\ee
The extra factor of 2 in front of the first term in (\ref{eq:V12})
in comparison with (\ref{eq:V13}) is due to the exchange 
effects in the case of same-flavour neutrino background. If the momentum 
distribution of the background neutrinos is isotropic, 
eq.~(\ref{eq:V13}) reduces to 
$\Delta V_\alpha=\pm \sqrt{2} G_F (N_{\nu_\beta}^L-N_{\bar{\nu}_\beta}^L)$.

Unlike in ordinary matter, neutrino potentials in neutrino backgrounds 
are not in general diagonal in the flavour basis. While the diagonal terms 
(\ref{eq:V12}) and (\ref{eq:V13}) arise from the coherent forward scattering 
processes 
$\nu_\alpha(\vec{k})+\nu_\beta(\vec{p})\to\nu_\alpha(\vec{k})+
\nu_\beta(\vec{p})$ 
(where the neutrino momenta are shown in the parentheses), the NC-induced 
momentum-exchange processes 
$\nu_\alpha(\vec{k})+\nu_\beta(\vec{p})\to\nu_\alpha(\vec{p})+
\nu_\beta(\vec{k})$ 
with $\alpha \ne \beta$ are also coherent and lead to flavour-off-diagonal 
potentials $V_{\alpha\beta}$ \cite{pantal1,pantal2,raffsigl,samuel}. 
The potential $V_{\alpha\beta}$ due to the scattering of a test neutrino 
of momentum $\vec{q}$ on background neutrinos and antineutrinos is  
\begin{align}
V_{\alpha\beta}=\sqrt{2} G_F \int\!&\frac{d^3p}{(2\pi)^3}\Big\{
\big(\rho_{\nu_\alpha \nu_\beta}^L(\vec{p})-\rho_{\bar{\nu}_\alpha 
\bar{\nu}_\beta}^L(\vec{p})\big)
(1-\cos\theta_{\vec{q}\vec{p}}) 
\nonumber \\[0.2em]
&-\frac{2E_{\nu_\alpha}(\vec{q})}{m_Z^2}\Big[E_{\nu_\beta}(\vec{p})
\rho_{\nu_\alpha \nu_\beta}^L(\vec{p})+E_{\bar{\nu}_\beta}(\vec{p})
\rho_{\bar{\nu}_\alpha \bar{\nu}_\beta}^L(\vec{p})\Big]
(1-\cos\theta_{\vec{q}\vec{p}})^2\Big\}.
\label{eq:V14}
\end{align}
Here $\rho_{\nu_\alpha \nu_\beta}^L(\vec{p})$ and $\rho_{\bar{\nu}_\alpha 
\bar{\nu}_\beta}^L(\vec{p})$ are the off-diagonal elements of the density 
matrices of left-handed neutrinos and their antiparticles in the flavour space:
\be
\rho_{\nu_\alpha \nu_\beta}^L(\vec{p})=\langle a^\dag_{\beta L}(\vec{p})
a_{\alpha L}(\vec{p})\rangle\,,
\label{eq:densmatrix}
\ee 
where $a^\dag_{\alpha L}(\vec{p})$ and $a_{\alpha L}(\vec{p})$ are the 
production and annihilation operators of $\nu_{\alpha L}(\vec{p})$, and 
similarly for antineutrinos. 
Note that the neutrino occupation numbers that enter in eqs.~(\ref{eq:V12}), 
(\ref{eq:occup}) and (\ref{eq:V13}) are the diagonal elements of these density 
matrices: $n_{\nu_\alpha}^L(\vec{p})=\rho_{\nu_\alpha \nu_\alpha}^L(\vec{p})$, 
$\,n_{\bar{\nu}_\alpha}^L(\vec{p})=\rho_{\bar{\nu}_\alpha 
\bar{\nu}_\alpha}^L(\vec{p})$. The off-diagonal potentials $V_{\alpha\beta}$ 
are in general complex, with $V_{\beta\alpha}=V_{\alpha\beta}^*$.   
Eq.~(\ref{eq:V14}) is valid for test neutrinos; for antineutrinos one has to 
replace 
\be
\rho_{\nu_\alpha \nu_\beta}^L(\vec{p})\leftrightarrow 
\rho_{\bar{\nu}_\alpha \bar{\nu}_\beta}^L(\vec{p})\,,\qquad 
E_{\nu_\alpha}(\vec{p})\to E_{\bar{\nu}_\alpha}(\vec{p})\,,\qquad 
E_{\nu_\beta}(\vec{p})\leftrightarrow E_{\bar{\nu}_\beta}(\vec{p})\,.
\vspace*{-0.5mm}
\label{eq:repl}
\ee

When considering neutrino flavour evolution in matter, one usually assumes 
that there is no back reaction of this evolution on the properties of the 
medium, and therefore matter-induced neutrino potentials are fixed external 
quantities. This is in general not true for neutrino oscillations in neutrino 
backgrounds, as the oscillations affect the state of the background. Therefore 
describing neutrino oscillations in media containing significant abundances of 
background neutrinos represents a complex non-linear problem. The elements of 
the neutrino and antineutrino density matrices in the flavour space that enter 
into eqs.~(\ref{eq:V12}), (\ref{eq:V13}) and (\ref{eq:V14}) must then be 
found self-consistently as solutions of the same flavour evolution problem. \\

\noindent
{\bf Magnetized matter}
\vspace*{1.5mm} \\
\noindent
In a medium with a magnetic field the particles of matter have in 
general non-zero average spin. In this case one can no longer neglect the 
axial-vector contributions to the neutrino potentials (see eq.~(\ref{eq:V2})). 
Under realistic conditions 
the average spin of the particles is relatively small, so that their 
polarizations are linear in the magnetic field strength. In this case in a 
matter consisting of electrons, protons and neutrons 
the neutrino potentials $V^0_\alpha$ get the extra contributions 
\begin{align}
&\Delta V^0_{\nu_e}=\pm(c_W^e+c_Z^e+c^p+c^n) B_{||}\,,\nonumber 
\\[0.3em]
&\Delta V^0_{\nu_\mu, \nu_\tau}=\pm(c_Z^e+c^p+c^n) B_{||}\,,
\label{eq:V11}
\end{align}
where $B_{||}$ is the component of the magnetic field along the neutrino 
velocity. The coefficients $c_W^e$ and $c_Z^e$ describe the contributions to 
the neutrino potentials coming from the polarization of the background 
electrons and caused by the CC and NC interactions respectively. The 
coefficients $c^p$ and $c^n$ are due to the polarization of the background 
protons and neutrons. 
For a relativistic gas of degenerate electrons (i.e., for $E_F\gg T$ 
where $E_F$ is the electron Fermi energy and $T$ is the temperature), such as 
e.g.\ in or near the supernova core, one has \cite{esposito,dolivo,elmfors}
\be
c_Z^e\simeq\frac{e G_F}{2\sqrt{2}}\left(\frac{3N_e}{\pi^4}\right)^{1/3}
\,,\quad c_W^e=-2c_Z^e\,.
\ee
For the contributions of the polarization of non-relativistic protons and 
neutrons with Boltzmann distributions functions one finds \cite{nunokawa,ALS2} 
\be
c^p\simeq \frac{G_F}{\sqrt{2}} g_A^p \frac{\mu_p \mu_N}{T}N_p\,,\qquad
c^n\simeq \frac{G_F}{\sqrt{2}} g_A^n \frac{\mu_n \mu_N}{T}N_n\,. 
\ee
Here $\mu_N=e/(2m_p)\simeq 3.152\times 10^{-18}$ MeV/G is the nuclear 
Bohr magneton, $\mu_p$ and $\mu_n$ are the proton and nucleon magnetic 
moments in units of the nuclear Bohr magneton ($\mu_p=2.793$, 
$\mu_n=-1.913$), and $g_A^p$ and $g_A^n$ are the NC axial-vector coupling 
constants of proton and neutron. 
For free nucleons, one has $g_A^p\simeq 1.36$ and $g_A^n\simeq -1.18$ 
\cite{raff,pdg}. In applications for neutron stars, the values of $g_A^p$ 
and $g_A^n$ in nuclear matter are more relevant; they can be estimated 
as free-space values divided by 1.27 \cite{raff}, i.e. $g_A^p\approx  
1.07$, $g_A^n\simeq -0.93$. Note that $c^p$, $c^n$ and $c_Z^e$ are all of 
the same sign. For non-degenerate particles, thermal fluctuations tend to 
destroy the polarization, and therefore $c^p$ and $c^n$ decrease with 
increasing temperature $T$.

\appsection
\renewcommand{\thesection}{\Alph{section}}
\section*{Appendix \Alph{section}: 
Proof of the equalities 
$\vec{p}_*=\vec{p}_K$ and $\vec{p}_*'=\vec{p}_{K'}$ }

We shall prove here that for macroscopic distances $|\vec{x}'-\vec{x}|$ 
the momentum integrals in the expression 
\be
\hat{S}_{\beta\alpha}(E;\vec{x}', \vec{x})=\int \frac{d^3 p}{(2\pi)^3}
\frac{d^3 p'}{(2\pi)^3}\,\tilde{S}_{\beta\alpha}(E;\vec{p}', \vec{p})\,
e^{i\vec{p}'\vec{x}'-i\vec{p}\vec{x}}\,, 
\label{eq:amp6a}
\ee
receive their main contributions from small regions around the points 
$\vec{p}=\vec{p}_*$ and $\vec{p}'=\vec{p}'_*$, which are defined 
as follows. For a given 
$E$, the value of $\vec{p}_*$ is obtained from the dispersion relation 
that stems from the neutrino evolution equation in matter of constant density 
equal to the density at the initial point of neutrino evolution $\vec{x}$. 
Likewise, $\vec{p}'_*$ is found from the neutrino 
dispersion relation in matter of constant density corresponding to the final 
point of neutrino evolution $\vec{x}'$.

Let us first consider $\tilde{S}_{\beta\alpha}(E;\vec{p}', \vec{p})$, which 
is a  Fourier transform of 
$\hat{S}_{\beta\alpha}(E;\vec{x}', \vec{x})$ (see eq.~(\ref{eq:S2})): 
\begin{align}
\tilde{S}_{\beta\alpha}(E;\vec{p}', \vec{p}) &=
\int d^3 x\, d^3 x'\,
\hat{S}_{\beta\alpha}(E;\vec{x}', \vec{x})
e^{-i\vec{p}'\vec{x}'+i\vec{p}\vec{x}}
\nonumber \\[0.3em]
&=-\frac{E}{2\pi}\int d^3 x\, d^3 x'\,
\frac{1}{|\vec{x}-\vec{x}'|}\hat{F}_{\beta\alpha}(E;\vec{x}', \vec{x})
e^{-i\vec{p}'\vec{x}'+i\vec{p}\vec{x}+i|E||\vec{x}-\vec{x}'|}\,. 
\label{eq:amp6b}
\end{align}
Here in the second line we used eq.~(\ref{eq:propF1}).  
For typical values of the energy $E$ and momenta $\vec{p}$ and $\vec{p}'$ 
of interest to us, the integrand in (\ref{eq:amp6b}) contains a fast 
oscillating phase factor, and therefore the integral can be calculated in the 
stationary phase approximation (see, e.g., \cite{erdelyi}). Defining  
\be
G(\vec{x}',\vec{x})\equiv -\vec{p}'\vec{x}'+\vec{p}\vec{x}
+ |E| |\vec{x}-\vec{x}'|
-i\ln  \hat{F}_{\beta\alpha}(E;\vec{x}', \vec{x})\,,
\label{eq:G}
\ee
we can rewrite eq.~(\ref{eq:amp6b}) as 
\be
\hat{S}_{\beta\alpha}(E;\vec{p}', \vec{p})=
-\frac{E}{2\pi}\int d^3 x\, d^3 x'\,
\frac{1}{|\vec{x}-\vec{x}'|}
e^{iG(\vec{x}',\vec{x})}\,. 
\label{eq:amp6c}
\ee
The main contributions to the integrals over the coordinates come from 
small neighbourhoods of the points where the phase $G(\vec{x}', 
\vec{x})$ is stationary. These points are found from the conditions 
\begin{align}
& \boldsymbol{\nabla}' G(\vec{x}', \vec{x})=-\vec{p}'
+i|E|\vec{\hat{r}}
-i \frac{\boldsymbol{\nabla}' \hat{F}_{\beta\alpha}(E; \vec{x}', \vec{x})}
{\hat{F}_{\beta\alpha}(E; \vec{x}', \vec{x})}=0\,,
\label{eq:stat1} \\[0.4em]
& \boldsymbol{\nabla} G(\vec{x}', \vec{x})=\vec{p}-|E|\vec{\hat{r}}
-i \frac{\boldsymbol{\nabla} \hat{F}_{\beta\alpha}(E; \vec{x}', \vec{x})}
{\hat{F}_{\beta\alpha}(E; \vec{x}, \vec{x}')}=0\,, 
\vspace*{-2mm}
\label{eq:stat2}
\end{align}
where $\boldsymbol{\nabla}'$ is the gradient with respect to the coordinate 
$\vec{x}'$ and $\vec{\hat{r}}\equiv (\vec{x}'-\vec{x})/|\vec{x}'-\vec{x}|$.  
Eqs.~(\ref{eq:stat1}) and (\ref{eq:stat2}) can also be rewritten 
\vspace*{-2mm} as 
\begin{align}
& i\boldsymbol{\nabla}' \hat{F}_{\beta\alpha}(E; \vec{x}', \vec{x})
=-(\vec{p}'-|E|\vec{\hat{r}})\,\hat{F}_{\beta\alpha}(E; \vec{x}', \vec{x})\,,
\label{eq:stat1a}   \\[0.4em]
& i \boldsymbol{\nabla} \hat{F}_{\beta\alpha}(E; \vec{x}', \vec{x})
=(\vec{p}-|E|\vec{\hat{r}})\,{\hat{F}_{\beta\alpha}(E; \vec{x}', \vec{x})}\,.
\label{eq:stat1b}
\end{align}
Eqs.~(\ref{eq:stat1a}), (\ref{eq:stat1b}) (or~(\ref{eq:stat1}), 
(\ref{eq:stat2})) should be solved with respect to the coordinates $\vec{x}$ 
and $\vec{x}'$ for fixed values of $\vec{p}$ and $\vec{p}'$. We denote the 
corresponding solutions $\vec{x}_*$ and $\vec{x}'_*$. Notice that $\vec{x}_*$ 
and $\vec{x}'_*$, 
are functions of $\vec{p}$ and $\vec{p}'$; we will not indicate this 
dependence explicitly in most of the following  
formulas in order not to overload the notation.

Applying the stationary phase approximation to 
eq.~(\ref{eq:amp6c}) yields
\be
\tilde{S}_{\beta\alpha}(E;\vec{p}', \vec{p})\approx 
e^{i\eta}\frac{E}{2\pi |\vec{x}_*-\vec{x}'_*|}
\sqrt{\frac{(2\pi)^6}{|D(\vec{x}_*',\vec{x}_*)|}}\;
e^{i G(\vec{x}_*',\vec{x}_*)}\,,
\label{eq:stphase1}
\ee
where $\eta$ is a constant phase which is of no relevance for us, and 
\be
D(\vec{x}_*',\vec{x}_*)\equiv \det 
\bigg[\bigg(
\frac{\partial^2 G(\vec{x}',\vec{x})}
{\partial \vec{x}_i' \partial \vec{x}_j}
\bigg)\!\Big|_{\vec{x}_*',\vec{x}_*}\, 
\bigg]\,.
\label{eq:stphase2}
\ee

Next, we substitute (\ref{eq:stphase1}) into (\ref{eq:amp6a}). 
Since for macroscopically separated $\vec{x}$ and $\vec{x}'$ the integrand of 
(\ref{eq:amp6a}) 
contains a fast oscillating phase factor (see section \ref{sec:simpl}), we can 
calculate the integrals over $\vec{p}$ and $\vec{p}'$ by once again making use 
of the stationary phase approximation.  
In doing so, we 
will need to find the stationary points of the expression 
\be
\tilde{G}(\vec{p}',\vec{p})\equiv \vec{p}'\vec{x}'-\vec{p}\vec{x}+  
G(\vec{x}'_*, \vec{x}_*)\,.
\label{eq:tildeG}
\ee
Here we have taken into account that $D(\vec{x}_*',\vec{x}_*)$ 
is not a fast oscillating function and therefore, in keeping with the 
stationary phase approximation, it need not be included in the phase 
factor $\tilde{G}(\vec{p},\vec{p}')$ but can instead be left as a 
pre-exponential factor. Substituting (\ref{eq:G}) into (\ref{eq:tildeG}) 
yields
\be
\tilde{G}(\vec{p}',\vec{p})\equiv \vec{p}'(\vec{x}'-\vec{x}'_*)
-\vec{p}(\vec{x}-\vec{x}_*)+|E||\vec{x}_*-\vec{x}'_*|
-i\ln  \hat{F}_{\beta\alpha}(E;\vec{x}'_*, \vec{x}_*)\,,
\label{eq:tildeG1}
\ee

Let us now find stationary points of $\tilde{G}(\vec{p}',\vec{p})$,  
which will give us the momenta that yield dominant contributions to the 
integrals over $\vec{p}$ and $\vec{p}'$ in (\ref{eq:amp6a}). 
Requiring that the derivatives of $\tilde{G}(\vec{p}',\vec{p})$ with respect to 
the components of $\vec{p}$ vanish, we find 
\begin{align}
0=\frac{\partial \tilde{G}(\vec{p}',\vec{p})}{\partial p_i}=
-(x&-x_*)_i -\frac{1}{\hat{F}_{\beta\alpha}(E; \vec{x}'_*, 
\vec{x}_*)}\bigg\{
\bigg[i\frac{\partial}{\partial {x}_{*j}}\hat{F}_{\beta\alpha}(E; \vec{x}'_*, 
\vec{x}_*) 
-(p_j-|E|\hat{r}_j)\hat{F}_{\beta\alpha}(E; \vec{x}_*', \vec{x}_*)\bigg]
\nonumber \\[0.2em]
\times & 
\frac{\partial x_{*j}}{\partial p_i}
+\bigg[i\frac{\partial}{\partial {x}'_{*j}}\hat{F}_{\beta\alpha}(E; \vec{x}'_*, 
\vec{x}_*)+(p'_j-|E|\hat{r}_j)\hat{F}_{\beta\alpha}(E; \vec{x}_*', \vec{x}_*)
\bigg]\frac{\partial x'_{*j}}{\partial p_i} \bigg\}.
\label{eq:stat3}
\end{align}
{}From eqs.~(\ref{eq:stat1a}) and (\ref{eq:stat1b}) it follows that the 
expressions in square brackets in (\ref{eq:stat3}) vanish, so that 
(\ref{eq:stat3}) simply reduces to $\vec{x}=\vec{x}_*(\vec{p},\vec{p}')$. 
Quite analogously, 
by requiring that the derivatives of $\tilde{G}(\vec{p}',\vec{p})$ with 
respect to the components of $\vec{p}'$ vanish, one finds 
$\vec{x}'=\vec{x}'_*(\vec{p},\vec{p}')$. Thus, the momenta at which the 
phase $\tilde{G}(\vec{p}',\vec{p})$ is stationary are obtained as the 
solutions of the system of equations
\begin{align}
& \vec{x}'=\vec{x}'_*(\vec{p},\vec{p}')\,,
\nonumber \\
& \vec{x}=\vec{x}_*(\vec{p},\vec{p}')\,.
\label{eq:system}
\end{align}
We will call the corresponding solutions $\vec{p}_*$ and $\vec{p}'_*$. 
Recall now that $\vec{x}_*$ and $\vec{x}'_*$ are the solutions of the 
system of equations (\ref{eq:stat1a}) and (\ref{eq:stat1b}) for fixed 
values of $\vec{p}$ and $\vec{p}'$. From eq.~(\ref{eq:system}) it follows 
that $\vec{p}_*$ and $\vec{p}'_*$ are the solutions of the same system 
(\ref{eq:stat1a}), (\ref{eq:stat1b}) which should now be considered as 
equations for the momenta 
at fixed values of the coordinates $\vec{x}$ and $\vec{x}'$. Note that when 
considered as equations for the momenta, eqs.~(\ref{eq:stat1a}) and 
(\ref{eq:stat1b}) are actually much simpler than when considered as equations 
for the coordinates; for known $\hat{F}_{\beta\alpha}(E;\vec{x}', \vec{x})$ 
one finds the solutions for the momenta $\vec{p}$ and $\vec{p}'$ immediately 
-- they are simply given by (\ref{eq:stat1}) and (\ref{eq:stat2}).
 
Next, we recall that the components of $\vec{p}_*$ and $\vec{p}'_*$ that are 
orthogonal to the vector $\vec{x}'-\vec{x}$ are negligibly small in all 
situations of practical interest (see discussion in section~\ref{sec:simpl}); 
therefore we are only interested in the longitudinal components of these  
momenta, which we denote ${\rm p}_*$ and ${\rm p}_*'$. Multiplying 
eqs.~(\ref{eq:stat1}) and (\ref{eq:stat2}) by $\vec{\hat{r}}F_{\beta\alpha}
(E;\vec{x},\vec{x}')$ yields 
\begin{align}
& i\frac{d}{dx'}F_{\beta\alpha}(E;\vec{x}',\vec{x})=-({\rm p}_*'-|E|)\, 
F_{\beta\alpha}(E;\vec{x}',\vec{x})\,,
\nonumber \\[0.4em]
& i\frac{d}{dx}F_{\beta\alpha}(E;\vec{x}',\vec{x})=
({\rm p}_*-|E|)\,F_{\beta\alpha}(E;\vec{x}',\vec{x})\,,
\label{eq:eq2}
\end{align}
where $d/dx'\equiv \vec{\hat{r}}\!\cdot\!\!\boldsymbol{\nabla}'$ and 
$d/dx \equiv \vec{\hat{r}}\!\cdot\!\boldsymbol{\nabla}$. On the other 
hand, we have 
\begin{align}
& i\frac{d}{dx'}\hat{F}(E;\vec{x}',\vec{x})=
H(\vec{x}')\,\hat{F}(E;\vec{x}',\vec{x})\,,
\label{eq:eq1a}
\\[0.4em]
& i\frac{d}{dx}\hat{F}(E;\vec{x}',\vec{x})=
-\hat{F}(E;\vec{x}',\vec{x})\,H(\vec{x})\,. 
\label{eq:eq2a}
\end{align}
where $H(\vec{x})=MM^\dag/2|E|+V(\vec{x})$. The first of these equations 
is just eq.~(\ref{eq:schroed0}), whereas the second one, which involves the 
differentiation of $\hat{F}(E;\vec{x}',\vec{x})$ with respect to the 
coordinate of the initial rather than final point of neutrino propagation,  
can be derived from the first one.%
\footnote{Indeed, $\hat{F}(E;\vec{x}',\vec{x})$ 
can be written as $\hat{F}(E;\vec{x}',\vec{x})=\hat{F}(E;\vec{x}',\vec{x}_1) 
\hat{F}(E;\vec{x}_1,\vec{x})$ with arbitrary $\vec{x}_1$. This relation 
can be easily verified by substituting it into eq.~(\ref{eq:eq1a}). 
Then, from $(d/dx_1)\hat{F}(E;\vec{x}',\vec{x})=0$ we have 
$[(d/dx_1)\hat{F}(E;\vec{x}',\vec{x}_1)]\hat{F}(E;\vec{x}_1,\vec{x})=
-\hat{F}(E;\vec{x}',\vec{x}_1)(d/dx_1)
\hat{F}(E;\vec{x}_1,\vec{x})$. 
Substituting here $(d/dx_1)\hat{F}(E;\vec{x}_1,\vec{x})$ from (\ref{eq:eq1a}) 
and multiplying the result by $[\hat{F}(E;\vec{x}_1,\vec{x})]^{-1}$ on the 
right, one arrives at (\ref{eq:eq2a}). 
}
Using eqs.~(\ref{eq:eq1a}) and (\ref{eq:eq2a}) in (\ref{eq:eq2}), we find
\begin{align}
 &({\rm p}_*'-|E|)\, \hat{F}(E;\vec{x}',\vec{x})=-H(\vec{x}')
\hat{F}(E;\vec{x}',\vec{x})
\label{eq:eq2c}
\\[0.4em]
& ({\rm p}_*-|E|)\,\hat{F}(E;\vec{x}',\vec{x})
=-\hat{F}(E;\vec{x}',\vec{x})H(\vec{x})\,.
\label{eq:eq2d}
\end{align}
Since the effective Hamiltonian $H(\vec{x})$ is non-diagonal in the 
flavour-eigenstate basis, eqs.~(\ref{eq:eq2c}) and (\ref{eq:eq2d}) are matrix 
equations for ${\rm p}_*$ and ${\rm p}'$. They are simplified in the local 
matter eigenstate bases defined in eqs.~(\ref{eq:mix2}) and (\ref{eq:diag}). 
In these bases the effective Hamiltonians ${\cal H}$ at the initial and final 
points of neutrino propagation are diagonal: 
${\cal H}(\vec{x})_{KM} = 
{\cal H}(\vec{x})_{K}\delta_{KM}$, ${\cal H}(\vec{x}')_{K'M'} = 
{\cal H}(\vec{x}')_{K'}\delta_{K'M'}$. Here ${\cal H}_K(\vec{z})$ is the $K$th 
local eigenvalue of $H$ at the point with the coordinate $\vec{z}$. Thus, we 
finally obtain from~(\ref{eq:eq2c}) and \vspace*{-3mm}(\ref{eq:eq2d}) 
\begin{align}
&{\rm p}'_* = {\rm p}'_{K'}\equiv|E|-{\cal H}_{K'}(\vec{x}')\,,
\label{eq:eq3a}
\\[0.4em]
&{\rm p}_*= {\rm p}_K\equiv|E|-{\cal H}_K(\vec{x})\,. 
\label{eq:eq3b}
\end{align}
Eqs.~(\ref{eq:eq3a}) and (\ref{eq:eq3b}) give the longitudinal (with respect 
to $\vec{x}'-\vec{x}$) components of the vectors $\vec{p}_*$ and $\vec{p}'_*$; 
as discussed above, their transverse components nearly vanish: 
\be
\vec{p}_{*\perp}\simeq 0\,,\qquad ~~\vec{p}'_{*\perp}\simeq 0\,.
\label{eq:eq3c}
\ee  
Note that eqs.~(\ref{eq:eq3a})-(\ref{eq:eq3c}) yield the correct neutrino 
dispersion relations in the limits of vanishing vacuum mixing or vanishing 
matter density.

Thus, we have proved that main contributions to the momentum integrals in 
(\ref{eq:amp6a}) come from small regions around the of momenta $\vec{p}_*$ 
and $\vec{p}_*'$, 
which satisfy the dispersion relations in matter at the initial and final 
points of neutrino propagation, respectively.

\appsection
\renewcommand{\thesection}{\Alph{section}}
\section*{Appendix \Alph{section}:\label{app:adiab}
Evolution equation in the adiabatic regime}

We shall prove here that in the adiabatic regime, when matter density varies 
sufficiently slowly along the neutrino path, the amplitude of the overall 
neutrino production-propagation-detection process (\ref{eq:amptotmatt}) 
satisfies the standard evolution equation (\ref{eq:Sch2}). 

In the adiabatic regime the transitions between different matter eigenstates 
are strongly suppressed, i.e.\ all matter eigenstates evolve independently. 
This means that the quantity $\hat{\cal F}$ that characterizes neutrino 
propagation in the matter eigenstate basis is diagonal: 
$\hat{\cal F}_{K'K}(E;\vec{x},\vec{x}_0) = 
\hat{\cal F}_{K}(E;\vec{x},\vec{x}_0)\delta_{K'K}$. The 
amplitude~(\ref{eq:amptotmatt})  can then be written as 
\be
{\cal A}_{\beta\alpha}^{\rm tot}(E,\vec{x},\vec{x}_0)=
\big\{\tilde{U}(\vec{x}) 
\big[\hat{\cal F}(E;\vec{x}, \vec{x}_0)^{}
\Phi_{P}\Phi_{D}\big]\tilde{U}^\dag (\vec{x}_0)
\big\}_{\beta\alpha}\,, 
\label{eq:ampl44}
\ee
where all the factors in the square brackets are diagonal.
{}From eq.~(\ref{eq:mattF}) we have 
\be
\hat{\cal F}(E;\vec{x}, \vec{x}_0)^{}=\tilde{U}^\dag(\vec{x}) 
\hat{F}(E;\vec{x},\vec{x}_0)\tilde{U}(\vec{x}_0)\,,
\label{eq:mattF2}
\ee
so that (\ref{eq:mattF2}) can be rewritten as 
\be
{\cal A}_{\beta\alpha}^{\rm tot}(E,\vec{x},\vec{x}_0)=\big\{
\hat{F}(E;\vec{x}, \vec{x}_0)^{}
\tilde{U}(\vec{x}_0) 
\Phi_{P}\Phi_{D}\tilde{U}^\dag (\vec{x}_0)\big\}_{\beta\alpha}\,, 
\label{eq:ampl45}
\ee
Substituting this into eq.~(\ref{eq:ampl44}) and differentiating, we obtain 
\begin{align}
i\frac{d}{dx}{\cal A}^{\rm tot}(E,\vec{x},\vec{x}_0)=&
i\frac{d}{dx}
\big\{
\hat{F}(E;\vec{x},\vec{x}_0)
\tilde{U}(\vec{x}_0) 
\Phi_{P}\Phi_{D}
\tilde{U}(\vec{x}_0)^\dag\big\} 
\nonumber \\[0.3em]
=&H(\vec{x})
\hat{F}(E;\vec{x},\vec{x}_0)
\tilde{U}(\vec{x}_0) 
\Phi_{P}\Phi_{D}\tilde{U}(\vec{x}_0)^\dag
=H(\vec{x}){\cal A}^{\rm tot}(E,\vec{x},\vec{x}_0)\,,
\label{eq:ampl46}
\end{align}
where we used eq.~(\ref{eq:schroed0}).

Thus, in the adiabatic regime the amplitude of the overall process satisfies 
the standard evolution equation (\ref{eq:Sch2}), irrespectively of whether or 
not the conditions of coherent neutrino production and detection are satisfied. 
However, the boundary condition for this amplitude differs from the standard 
one. 
Instead, from eqs.~(\ref{eq:ampl45}) and (\ref{eq:boundary}) we find  
\be
{\cal A}^{\rm tot}_{\beta\alpha}(E,\vec{x},\vec{x}_0)|_{\vec{x}\to 
\vec{x}_0}=
\sum_K \tilde{U}(\vec{x}_0)_{\alpha K}^*\tilde{U}^{}_{\beta K}(\vec{x}_0)
\Phi_{P}(E, {\rm p}_{K})\Phi_{D}(E,{\rm p'}_K) 
=\big\{\tilde{U}(\vec{x}_0) \Phi_{P}\Phi_{D}\tilde{U}^\dag (\vec{x}_0)
\big\}_{\beta\alpha}\,.
\label{eq:boundary2}
\ee
If the coherence conditions for neutrino production and detection 
(\ref{eq:Condit1}), (\ref{eq:Condit2}) are satisfied, one can replace 
the momenta ${\rm p}_K$ and ${\rm p'}_K$ in the arguments of the 
amplitudes $\Phi_P$ and $\Phi_D$ in eq.~(\ref{eq:boundary2})
by the corresponding average values and pull these amplitudes from the 
sum. Eq.~(\ref{eq:boundary2}) then reduces, up to a constant factor, to the 
standard boundary condition: ${\cal A}^{\rm tot}_{\beta\alpha}
(E,\vec{x},\vec{x}_0)|_{\vec{x}\to \vec{x}_0}=\delta_{\beta\alpha} 
\Phi_P(E,{\rm p}) \Phi_D(E,{\rm p}')$. 


\end{document}